\documentclass[twocolumn,twoside,showkeys,showpacs,preprintnumbers,superscriptaddress,amsmath,amssymb,prl]{revtex4-2}
\usepackage{graphicx}% Include figure files
\usepackage{dcolumn}% Align table columns on decimal point
\usepackage{bm}% bold math
\usepackage[version=4]{mhchem}
\usepackage{subcaption}
\usepackage[colorlinks=true,linkcolor=black,citecolor=blue,urlcolor=blue,]{hyperref}
\usepackage{appendix}
\usepackage{ulem} 
\usepackage[justification=raggedright,singlelinecheck=false]{caption}

\usepackage{pstricks}
\usepackage{amsmath, amsthm, amssymb}
\usepackage{changes}

\def\snsn{ $\rm ^{124}Sn$+$\rm ^{124}Sn$} 
\def\rhmt1{}
\def\rhmt2{}

\def\ssdtwo {\texttt{SSD M2}}
\def\ssdgam {\texttt{SSD M1 \& CsI M1}}
\def\ssdneutron {\texttt{SSD M1 \& NA M1}}
\def\ssdls {\texttt{SSD M1 \& LS}}
\def\lstzero {\texttt{LS \& T$_0$}}
\def\ntzero {\texttt{NA M1 \& T$_0$}}

\begin{document}

\title{Precise Measurement of Short-Range Correlations in Nuclei from Bremsstrahlung Gamma Ray Emission in Low-Energy Heavy-Ion Collisions}

\author{Junhuai Xu}
\email{xjh22@mails.tsinghua.edu.cn}
\affiliation{Department of Physics, Tsinghua University, Beijing 100084, China}%

\author{Qinglin Niu}
\affiliation {School of Physics, Nanjing University, Nanjing 210093, China}

\author{Yuhao Qin}%
\email{qinyh18@tsinghua.org.cn}
\affiliation{Department of Physics, Tsinghua University, Beijing 100084, China}%

\author{Dawei Si}
\affiliation{Department of Physics, Tsinghua University, Beijing 100084, China}%

\author{Yijie Wang}
\affiliation{Department of Physics, Tsinghua University, Beijing 100084, China}%

\author{Sheng Xiao}
\affiliation{Department of Physics, Tsinghua University, Beijing 100084, China}%

\author{Baiting Tian}
\affiliation{Department of Physics, Tsinghua University, Beijing 100084, China}%

\author{Zhi Qin}
\affiliation{Department of Physics, Tsinghua University, Beijing 100084, China}%

\author{Haojie Zhang}
\affiliation{Department of Physics, Tsinghua University, Beijing 100084, China}%

\author{Boyuan Zhang}
\affiliation{Department of Physics, Tsinghua University, Beijing 100084, China}%

\author{Dong Guo}
\affiliation{Department of Physics, Tsinghua University, Beijing 100084, China}%

\author{Minxue Fu}
\affiliation{Department of Physics, Tsinghua University, Beijing 100084, China}%

\author{Xiaobao Wei}
\affiliation {Institute of Particle and Nuclear Physics, Henan Normal University, Xinxiang 453007, China}

\author{Yibo Hao}
\affiliation {Institute of Particle and Nuclear Physics, Henan Normal University, Xinxiang 453007, China}

\author{Zengxiang Wang}
\affiliation {Institute of Particle and Nuclear Physics, Henan Normal University, Xinxiang 453007, China}

\author{Tianren Zhuo}
\affiliation {Institute of Particle and Nuclear Physics, Henan Normal University, Xinxiang 453007, China}

\author{Chunwang Ma}
\affiliation {Institute of Particle and Nuclear Physics, Henan Normal University, Xinxiang 453007, China}
\affiliation {Institute of Nuclear Science and Technology, Henan Academy of Sciences, Zhengzhou, 450015, China}

\author{Yuansheng Yang}
\affiliation {Institute of Modern Physics, Chinese Academy of Sciences, Lanzhou 730000, China}

\author{Xianglun Wei}
\affiliation {Institute of Modern Physics, Chinese Academy of Sciences, Lanzhou 730000, China}

\author{Herun Yang}
\affiliation {Institute of Modern Physics, Chinese Academy of Sciences, Lanzhou 730000, China}

\author{Peng Ma}
\affiliation {Institute of Modern Physics, Chinese Academy of Sciences, Lanzhou 730000, China}

\author{Limin Duan}
\affiliation {Institute of Modern Physics, Chinese Academy of Sciences, Lanzhou 730000, China}

\author{Fangfang Duan}
\affiliation {Institute of Modern Physics, Chinese Academy of Sciences, Lanzhou 730000, China}

\author{Kang Wang}
\affiliation {Institute of Modern Physics, Chinese Academy of Sciences, Lanzhou 730000, China}

\author{Junbing Ma}
\affiliation {Institute of Modern Physics, Chinese Academy of Sciences, Lanzhou 730000, China}

\author{Shiwei Xu}
\affiliation {Institute of Modern Physics, Chinese Academy of Sciences, Lanzhou 730000, China}

\author{Zhen Bai}
\affiliation {Institute of Modern Physics, Chinese Academy of Sciences, Lanzhou 730000, China}

\author{Guo Yang}
\affiliation {Institute of Modern Physics, Chinese Academy of Sciences, Lanzhou 730000, China}

\author{Yanyun Yang}
\affiliation {Institute of Modern Physics, Chinese Academy of Sciences, Lanzhou 730000, China}

\author{Mengke Xu}%
\affiliation {Shanghai Advanced Research Institute, Chinese Academy of Sciences, Shanghai 201210, China}
\author{Kaijie Chen}%
\affiliation {Shanghai Advanced Research Institute, Chinese Academy of Sciences, Shanghai 201210, China}
\author{Zirui Hao}%
\affiliation {Shanghai Advanced Research Institute, Chinese Academy of Sciences, Shanghai 201210, China}
\author{Gongtao Fan}%
\affiliation {Shanghai Advanced Research Institute, Chinese Academy of Sciences, Shanghai 201210, China}
\author{Hongwei Wang}%
\affiliation {Shanghai Advanced Research Institute, Chinese Academy of Sciences, Shanghai 201210, China}

\author{Chang Xu}
\email{cxu@nju.edu.cn}
\affiliation {School of Physics, Nanjing University, Nanjing 210093, China}

\author{Zhigang Xiao}
\email{xiaozg@tsinghua.edu.cn}
\affiliation{Department of Physics, Tsinghua University, Beijing 100084, China}%
\affiliation{Center of High Energy Physics, Tsinghua University, Beijing 100084, China}

\date{\today}

\begin{abstract}

Atomic nuclei and dense nucleonic matter in neutron stars exhibit short-range correlations (SRCs), where nucleons form temporally correlated pairs in proximity beyond mean-field approximation. It is essential to make precision measurement of the fraction of SRC since it carries the signature of underlying quark dynamics in nuclear medium.
In this letter, we present the first high-precision measurement of neutron-proton bremsstrahlung $\gamma$-ray emission from the symmetric \snsn~ reactions at 25 MeV/u. From the observed spectral hardening, the precise SRC fraction in the $\rm ^{124}Sn$ nucleus is extracted to be $(20 \pm 3)\%$. 
This result provides a novel, direct and unambiguous evidence of SRCs, and demonstrates that low-energy heavy-ion collisions offers a new approach to studying nuclear structure in connection with quark-level dynamics.

\end{abstract}

\maketitle

{\it Introduction $-$} Nucleonic matter, composed of strongly interacting protons and neutrons, exhibits complex many-body dynamics. Among these, some nucleons form short-range correlations (SRCs) \cite{Hen:2014nza, Hen:2016kwk}, which are temporal fluctuation effects that go beyond the mean-field approximation. SRCs have been experimentally observed and provide an explanation for the high-momentum tail (HMT) observed beyond the Fermi surface in nucleon momentum distributions \cite{Hen:2014nza, CLAS:2018yvt}. These high-momentum components significantly influence the properties of nuclear matter, particularly the density dependence of the nuclear symmetry energy \cite{PhysRevC.81.064612, PhysRevC.91.025803, PhysRevC.93.014619}, a quantity that is crucial in dense environments like neutron stars and heavy-ion collisions (HICs). SRCs reflect the short-range structure of the nuclear force, and their precise measurement is key to uncovering the origin of nucleon-nucleon interactions.

Studies indicate that SRCs are primarily characterized by neutron-proton ($np$) pairs in the isosinglet (T=0) channel at short distances, favoring $ud$ diquark isosinglet and spin-singlet configurations \cite{RITTENHOUSEWEST2023122563}. These structural features of SRCs have been shown to significantly influence the EMC effect \cite{EuropeanMuon:1983wih} in nuclei \cite{Weinstein:2010rt, Hen:2012fm, CLAS:2019vsb}, establishing a direct link between short-range nucleon correlations and quark-level modifications of nuclear structure \cite{Hen:2016kwk, Norton:2003cb, Geesaman:1995yd, Malace:2014uea, Kelly:1996hd, Dickhoff:2004xx}. Furthermore, a recent global analysis based on the QCD factorization scheme has extracted the universal effective parton distribution within SRC pairs and the mean-field fraction using a variety of high-energy data \cite{PhysRevLett.133.152502}. These findings suggest that SRC studies may offer valuable insights into the modification of quark dynamics in the nuclear environment, serving as a potential bridge between nuclear structure and the underlying partonic degrees of freedom.

The majority of our present understanding of SRC  comes from the electron-nucleus scattering \cite{JeffersonLabHallA:2007lly,Subedi:2008zz,LabHallA:2014wqo,Hen:2014nza,CLAS:2018yvt} and proton-nucleus knockout \cite{Tang:2002ww,Piasetzky:2006ai,Patsyuk2021} experiment.   Among various experimental approaches, HICs offer a unique opportunity to create bulk dense nucleonic matter in  terrestrial laboratory. Therefore, searching for signatures of SRC in HICs is of importance, as it not only links SRC to dense nucleonic matter, but also opens  new avenues for quantifying the SRC fraction under extreme conditions.  When two heavy nuclei collide, two-body  $np$ collisions generate bremsstrahlung $\gamma$-rays \cite{PhysRevC.34.2127,MARTINEZ199928,Gan1994298}. Due to the influence of the high momentum tail (HMT) arising from SRCs, the average nucleon kinetic energies of both projectile and target increase significantly, resulting in a hardening of the emitted $\gamma$-ray spectrum, particularly at high energies \cite{Xue:2016udl}. In comparison  to the hadron probes \cite{PhysRevC.96.054603,Hagel.WPCF2023,Huang:2025uvc}, the high-energy $\gamma$-rays are largely unaffected by the nuclear medium, making them a clean and unambiguous probe for detecting SRC effects \cite{Xue:2016udl,Guo:2021zcs,Xu:2012hf}.

Theoretical calculations using the Isospin-dependent Boltzmann-Uehling-Uhlenbeck (IBUU) transport model \cite{BERTSCH1988189}, incorporating
Momentum-Dependent Interaction (MDI) \cite{PhysRevC.67.034611}, predict the bremsstrahlung $\gamma$-ray spectra produced in HICs by varying the HMT fraction ($R_{\text{HMT}}$), which quantifies the proportion  of nucleons with momenta above the Fermi surface in nuclei \cite{Xue:2016udl}. By comparing model predictions with experimental $\gamma$-ray spectrum, The HMT value $R_{\text{HMT}}$ can be effectively constrained. In our previous experiment involving $\rm ^{86}Kr$+$\rm ^{124}Sn$ collisions, the bremsstrahlung $\gamma$-ray spectrum was  measured, delivering a nonzero HMT fraction induced by SRCs \cite{Qin:2023qcn,Xu:2024oct}. The results showcased the potential of bremsstrahlung $\gamma$-ray emission as a novel probe of SRC in nuclei.

This letter reports on a new and high precision  measurement of bremsstrahlung $\gamma$-rays from \snsn~ collisions at 25 MeV/u with significantly  enhanced statistics and extended high energy end of the spectrum. The $\gamma$-ray spectrum is analyzed using the IBUU-MDI transport model, which incorporates the $np\rightarrow np\gamma$ process.  The use of a symmetric reaction allows for an unambiguous extraction of $R_{\text{HMT}}$ in the $\rm ^{124}Sn$ nucleus.

\begin{figure}[hbtp]
    \centering
    \includegraphics[width=0.45\textwidth]{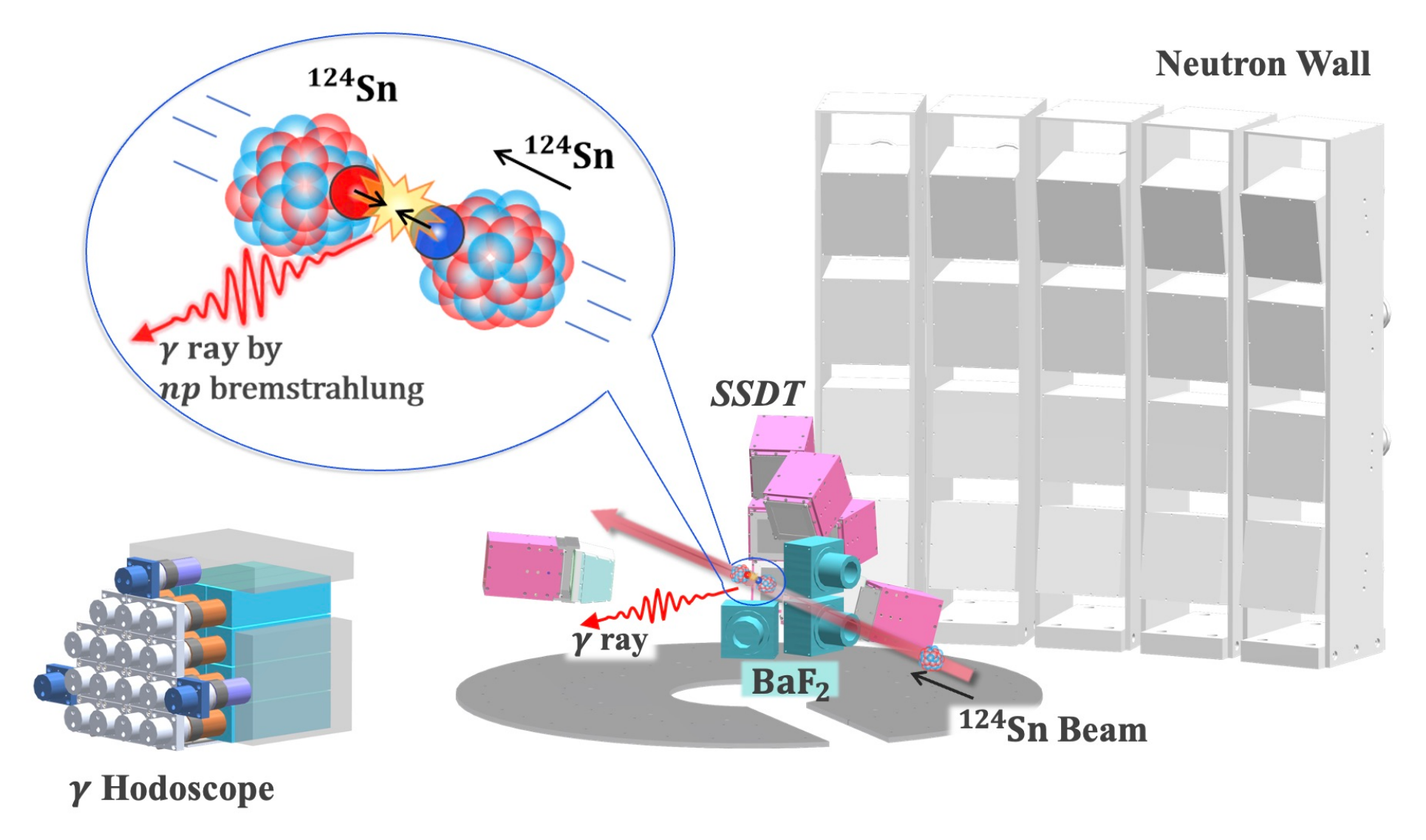}
    \caption{An illustration of the CSHINE detector setup. A schematic view of an \snsn~ event is plotted, involving the $np$ collision that emits a bremsstrahlung $\gamma$-ray, subsequently detected by the CsI(Tl) hodoscope with three surrounding plastic scintillators.}
    \label{detectorsetup}
\end{figure}

{\it Experiment$-$}  The experiment was performed on Radioactive Ion Beam Line 1 (RIBLL1) at the Heavy Ion Research Facility in Lanzhou (HIRFL). The products from the reaction of 25 MeV/u \snsn~  were measured by the Compact Spectrometer for Heavy Ion Experiment (CSHINE) \cite{Si:2024ujh, Guan:2021tbi, Wang:2021jgu}. Fig. \ref{detectorsetup} presents the detector setup. The charged isotopes of $Z\le 5$ were identified and reconstructed by 6  silicon-strip detector telescopes (SSDTs) covering  $20^\circ < \theta_{\rm lab} <60^{\circ}$ in partial azimuth \cite{Guan:2021nfk,NuclSciTech.36.132}. Neutrons are  detected by a $4\times5$ array  of plastic scintillator installed at 2.0 meters to the target. With 4 fast $\rm BaF_2$ $\gamma$-detectors surrounding the target delivering starting time (T$_0$) information, neutrons can be discriminated from the $\gamma$-rays \cite{Si:2024ujh,Si:2025eou}. Besides, a liquid scintillator (LS) neutron detector of $\rm \phi 348~mm\times1020~mm$ is mounted at 513 cm  to the target (not shown in Fig. \ref{detectorsetup}).

To measure the energetic bremsstrahlung  $\gamma$-rays from the reactions on the target, a $4 \times 4$ CsI(Tl)-based hodoscope has been mounted \cite{Qin:2022mzp}. Each unit has a dimension of $70 \times 70 \times 250~\rm mm^3$. Three plastic scintillators were installed surrounding the CsI(Tl) hodoscope to veto the cosmic ray background as well as the $\gamma$ events with energy leaking to the outside. A dual-range amplifier was applied to record the small and large energy signals, and then the response of each CsI(Tl) unit was calibrated by a dual-range method \cite{Xu:2025erv}.  The energy range of each individual CsI(Tl) unit was optimized in order to reach the maximum $\gamma$-ray energy deposit of 100 MeV in the hodoscope.  The background, which mainly includes cosmic-ray muons and residual $\gamma$ radiation at low energies,  was measured when the beam was off.  

The data acquisition was triggered by one of six conditions listed below.
$\rm Trg_1$, one CsI (Tl) unit and one SSDT fired (\ssdgam); 
$\rm Trg_2$, two SSDT fired (\ssdtwo); 
$\rm Trg_3$, one neutron array unit and one SSDT fired (\ssdneutron); 
$\rm Trg_4$, liquid scintillator and one SSDT fired (\ssdls);
$\rm Trg_5$, one neutron array unit and one T$_0$ fired (\ntzero);
$\rm Trg_6$, liquid scintillator and one T$_0$ fired (\lstzero).
The trigger scheme was implemented using field-programmable gate array (FPGA) technology \cite{Guo:2022kwc}.

\begin{figure}[hbtp]
    \centering
    \includegraphics[width=0.45\textwidth]{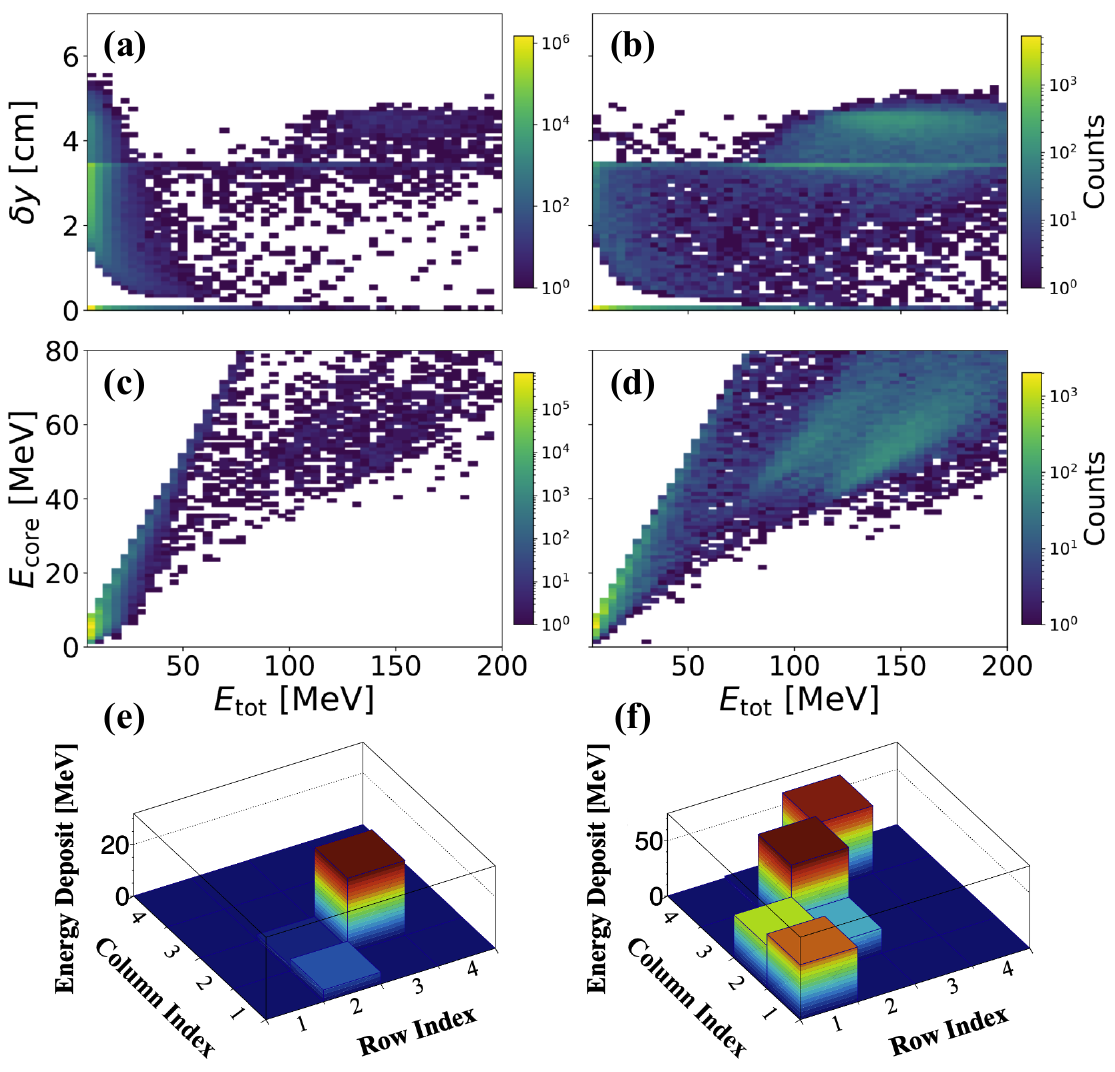}
    \caption{Event-wise vertical spreading $\delta_y$  as a function of the total reconstructed energy $E_{\rm tot}$ for beam-on (a)  and beam-off (b)  data. Correlation between the energy deposited in the event core  $E_{\rm core}$ and  $E_{\rm tot}$ for  beam-on (c)  and beam-off (d) data. A typical  display of a high energy $\gamma$ event from the reaction (e), and  of a cosmic muon penetration event through the $\gamma$-hodoscope (f).}
    \label{BeamandBKG}
\end{figure}

After calibrating the CsI(Tl) hodoscope, the energy deposits are reconstructed event-by-event. In each triggered event, the reconstruction algorithm identifies the clustered CsI(Tl) hits coincided in a $|\Delta T|<50$ ns time window and led by a `{\it core}' with the largest energy deposit.  The $\gamma$ events and the penetrating cosmic muon events can be distinguished by their topological characteristics, as depicted in Fig.~\ref{BeamandBKG} in beam-on (left panels) and beam-off (right panels) stages, respectively. For the beam-on measurement, on the scattering plots of  the variance  of the spatial distribution of the fired  units $\delta_y$  \cite{Qin:2022mzp} (a) and the energy deposit in the core $E_{\rm core}$  (c) as a function of the total reconstructed energy $E_{\rm tot}$, two dominant components are clearly visible. One is the $\gamma$ event of small spatial extension and in the vicinity of $E_{\rm core} \approx E_{\rm tot}$, indicating that most of the energy is collected by the event core with less extension to the neighbouring units. The other is dominated by the cosmic muon characterized by large $\delta_y$ and $ E_{\rm tot}>E_{\rm core}$, indicating the incident muon fires several CsI(Tl) units with approximate energy deposit. The events with large $E_{\rm tot}$ in the beam-on data exhibit similarity with those observed in the beam-off background, as shown in (b) and (d), where cosmic muons dominate in large $E_{\rm tot}$ area. Fig.~\ref{BeamandBKG}(e) and (f) illustrate a typical  $\gamma$-ray event and a penetrating cosmic muon event, respectively.  Low-energy events ($E_{\rm tot} < 30$ MeV) with multiple hits mainly result from statistical residual $\gamma$ emissions. 

\begin{figure}[hbtp]
    \centering
    \includegraphics[width=0.45\textwidth]{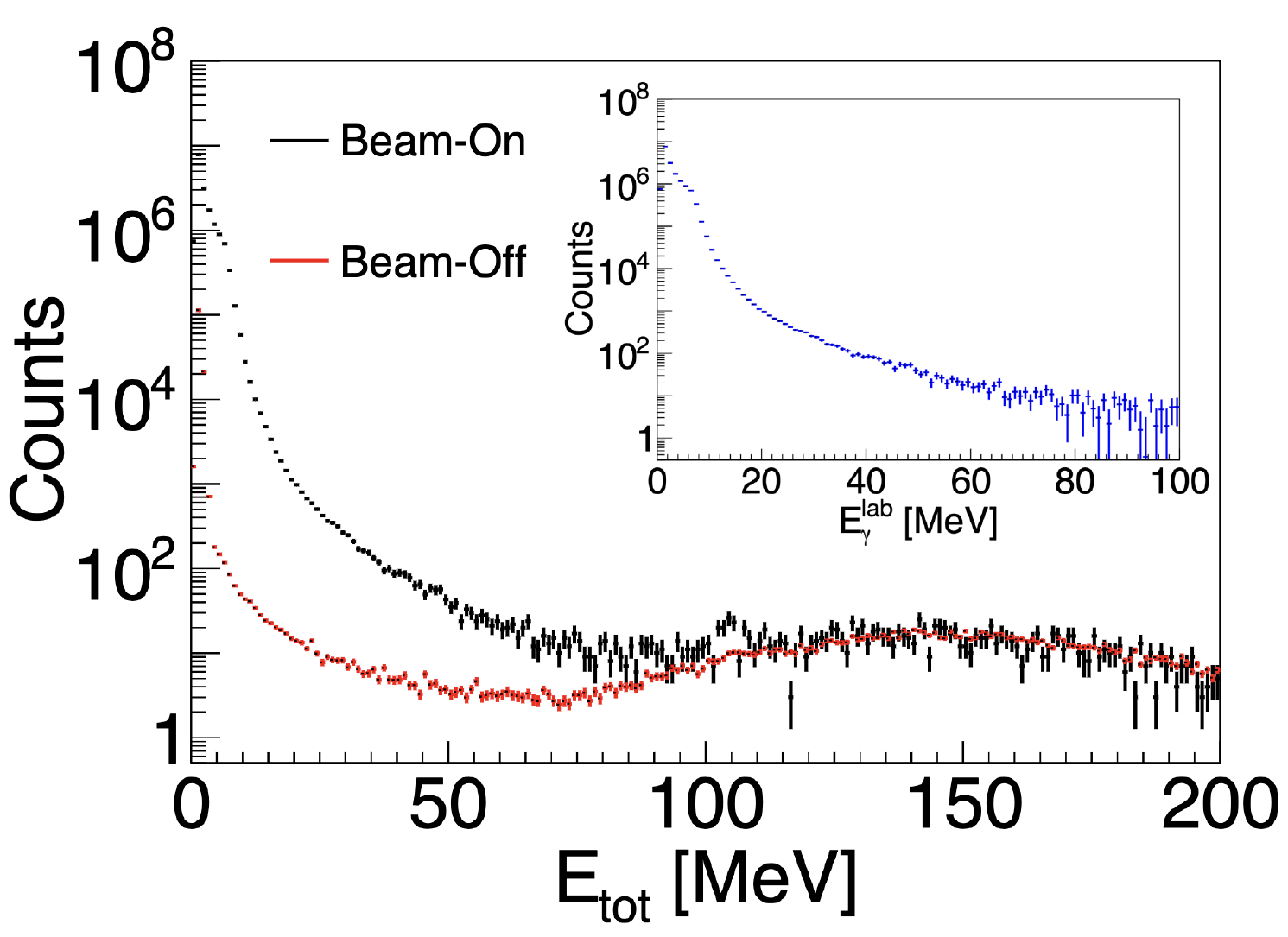}
    \caption{The experimental $\gamma$ energy spectrum in laboratory  measured with beam-on (black), compared to the background spectrum with beam-off (red), normalized to the counts above $110~\rm MeV$. The inset displays the background-subtracted $\gamma$ spectrum in laboratory.}
    \label{BKGSubtraction}
\end{figure}

Fig.~\ref{BKGSubtraction} presents the total  energy  spectra in beam-on (black) and beam-off (red) measurements. No selection on the impact parameter $b$ is done, since the spectrum shape shows insignificant dependence on $b$ \cite{Qin:2023qcn}. The beam-off background  is scaled to match the beam-on spectrum using the counts above $110~\rm MeV$. The pure  $\gamma$  ray spectrum in laboratory  can be  obtained by subtracting the scaled background spectrum from the total spectrum, as displayed in the inset. It is noted here that the spectrum is affected by detector efficiency, and hence the theoretical spectra must be filtered by the same detector response before comparison can be made (see next). The detector filter matrix, describing the response of the CsI(Tl) hodoscope  to $\gamma$-rays of varying energies in laboratory, was generated using Geant4 simulations  \cite{Qin:2022mzp}.

The main systematic uncertainty originates from detector calibration with the linear response assumption, as the radioactive sources  cover energies below 3 MeV, whereas the bremsstrahlung $\gamma$-rays may deposit tens of MeV in a single CsI(Tl) unit. Higher energy $\gamma$-ray beam test was conducted to characterize the nonlinearity of the CsI(Tl) unit. As a result, the detector response correction functions (DRCFs) were obtained to quantify the deviations from linear calibration. The uncertainty brought by the linear calibration is within 4\% \cite{Xu:2025erv}.

{\it Results and Discussions $-$} The IBUU transport model \cite{Li:1996ix,Li:1997rc,Li:2003ts} simulates nucleus-nucleus collisions and has been extended to include isospin degrees of freedom \cite{Li:2018lpy} and momentum-dependent nucleon interaction potential (MDI) \cite{Das:2002fr}. The SRC and HMT effects are introduced in the initialization of single-nucleon momentum distribution $\rho(k)$ \cite{LI201829,Xue:2016udl}.  For the asymmetric nuclear matter (ANM), the high-momentum component depends linearly  on the isospin asymmetry $\delta$ \cite{PhysRevC.79.064308,PhysRevC.87.014314,PhysRevC.93.014619}. Thus $\rho(k)$ for ANM is parameterized  as
\begin{equation}
\rho_{\rm N}(k)=\left\{
\begin{array}{rcl}
B_{\rm N}, & & {k\le k^{\rm N}_{\rm F}}\\
\eta(1+s\delta)/k^4, & & {k^{\rm N}_{\rm F}<k\le \lambda k^{\rm N}_{\rm F}}\\
0, & & {k>\lambda k^{\rm N}_{\rm F}}
\end{array} \right.
\end{equation}
where the subscript or superscript $\rm N$ denotes either a proton ($s = +1$) or a neutron ($s = -1$), $k$ is the single-nucleon momentum, $k_{\rm F}$ is the Fermi momentum and $\lambda$ is the high-momentum cutoff \cite{Xue:2016udl}. Given the HMT fraction $R_{\rm HMT}$, the  constants $B_{\rm N}$ and $\eta$ are determined by $4\pi \int\limits_{0}^{\infty}\rho_{\rm N}(k)k^{2}dk=1$ and $4\pi \int\limits_{k_{\rm F}}^{\infty}\rho_{\rm N}(k)k^{2}dk=R_{\rm HMT}$, respectively. The bremsstrahlung $\gamma$ production channel is incorporated in the $N$-$N$ collision processes governed by the Bertsch criterion \cite{TMEP:2022xjg} and Pauli exclusion principle. 

Due to the low production rate of the bremsstrahlung $\gamma$, its impact on the nucleon kinematics is negligible. Perturbation method is then employed to calculate the photon production probability in each $np$ collision. Based on the one boson exchange (OBE) model, a well fitted formula is applied to calculate the elementary double differential $\gamma$ production probability  \cite{Gan1994298},
\begin{equation}
    \frac{d^2P}{d\Omega dE_{\gamma}} = 1.67\times 10^{-7}\times\frac{[1-(E_{\gamma}/E_{\rm max})^2]^{\alpha}}{E_{\gamma}/E_{\rm max}}.
\end{equation}
Here, $E_{\gamma}$ represents the energy of produced photons, and $E_{\rm max}$ is the total available energy in the center-of-mass (CM) frame of the colliding $np$ pair. The coefficient $\alpha$ reads $\alpha = 0.7319 - 0.5898\beta_i$, where $\beta_i$ is the nucleon velocity. The total $\gamma$ yield per event is obtained by summing over all photon-producing $np$ collisions during the reaction. The  $\gamma$ spectrum show minimal dependence on the nuclear mean-field, impact parameter, and symmetry energy in the test,  as expected because bremsstrahlung photons are predominantly emitted in the early stages and interact only electromagnetically.

\begin{figure}[hbtp]
    \centering
    \includegraphics[width=0.45\textwidth]{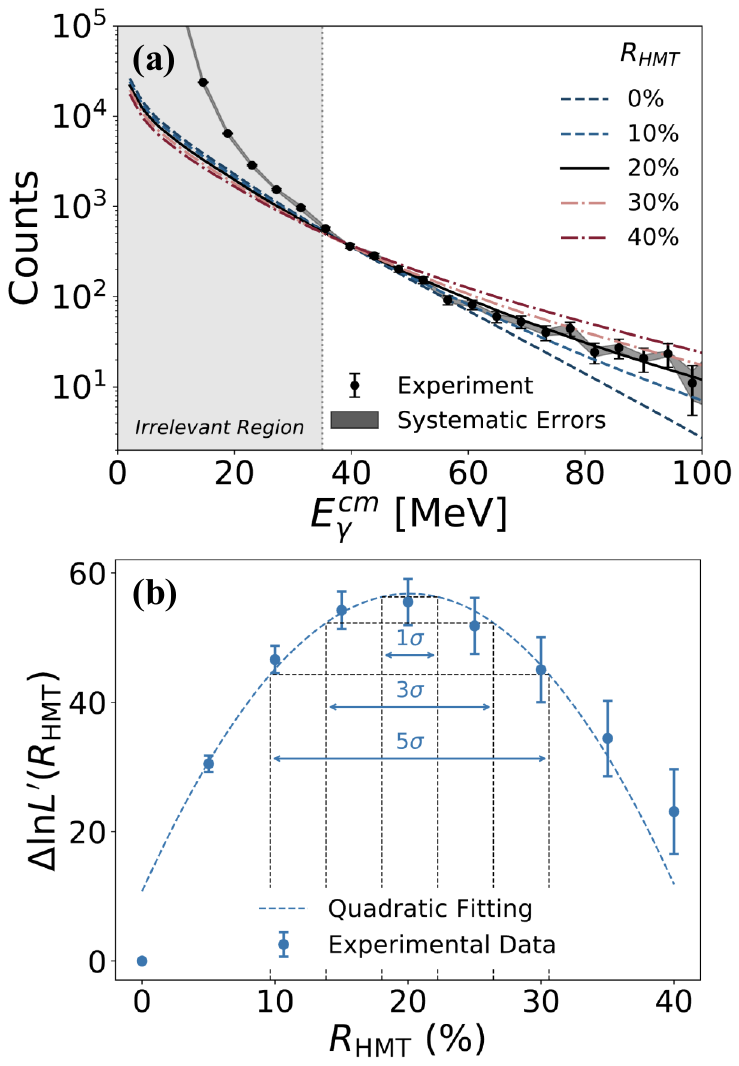}
    \caption{ (a) Comparison of the rebinned experimental $\gamma$ spectrum (black dots) in CM reference with various detector-filtered theoretical curves. Statistical (systematic) uncertainties are denoted by vertical bars (gray bands). (b) Likelihood distribution as a function of $R_{\rm HMT}$ with quadratic fit (dashed curve). The error bars represent the standard deviation of likelihood values by varying the calibration and detector response data sets. Several main confidence levels are indicated.}
    \label{RHMTFtting}
\end{figure}

Fig.~\ref{RHMTFtting} (a) illustrates the rebinned experimental  $\gamma$  spectrum converted to the CM frame. The black dots indicate the central values with statistical errors, while gray bands represent systematic uncertainties. Several key theoretical curves, filtered by the detector response in laboratory, are presented for comparison. Unlike our previous experiment \cite{Qin:2023qcn, Xu:2024oct}, the  $\gamma$  detector array was operated in active triggering mode, preventing the determination of the number of $np$ collisions in HICs for normalizing the $\gamma$ spectrum. Therefore, the comparisons with model predictions focus on spectral shape rather than absolute yield. Theoretical spectra were uniformly scaled by $2.5 \times 10^9$ to match the data range, without affecting $R_{\rm HMT}$ extraction based on shape analysis.

A maximum likelihood analysis was performed to compare the  model prediction with the experimental data, using the likelihood function \cite{Qin:2023qcn} defined as

\begin{equation}
    \ln L'(R_{\text{HMT}}) = \sum_{i}^{\text{range}}n_i\ln p_i(R_{\text{HMT}}).
\label{rlhood}    
\end{equation}
where $i$  sums over the experimental points within the statistical analysis interval, and $n_i$ denotes the count in the $i^{\rm th}$ experimental bin. The quantity $p_i$  represents the probability that the theoretical model predictions fall into the corresponding histogram bin  under the specified statistical analysis interval for a given  $R_{\text{HMT}}$. $\gamma$ rays below $30~\rm MeV$ are dominated by collective resonances, statistical emissions and possibly residual radiation in the target chamber. Accordingly, the analysis was restricted to the range of interest $35 < E_{\gamma}^{\rm cm} < 100~\rm MeV$ to evaluate the likelihood between experimental spectra and IBUU-MDI predictions for different $R_{\rm HMT}$ values.

The results are shown as blue dots with error bars in Fig.~\ref{RHMTFtting}(b). To standardize the likelihood function values obtained from different $\gamma$  spectra, the likelihood corresponding to the theoretical curve with $R_{\text{HMT}} = 0\%$  was selected as the reference point ($\ln L'(R_{\text{HMT}} = 0\%) = 0$). The relative likelihood values, $\Delta\ln L'(R_{\text{HMT}})$, for all other points were calculated relative to this baseline.   A quadratic fit over $R_{\rm HMT}=5–35\%$ is conducted as shown by the blue dashed curve. The optimal value  $R_{\text{HMT}} = (20.1 \pm 2.1)\%$ is extracted. Considering the variation brought by fitting procedure and by using different scheme of background subtraction, the final constraint writes $R_{\text{HMT}} = (20 \pm 3)\%$. As a test of the look-else-effect, the optimal $R_{\rm HMT}$ value is unchanged against the variation of the low limit of $E_{\gamma}^{\rm cm}$ from 33 to 44 MeV. 

The result reveals a potential new paradigm to probe the PDF of nucleon in nuclei (nPDF) via low energy HICs. According to the collinear factorization scheme \cite{nCTEQ:2023cpo}, the nPDF parametrization is composed of a linear combination of PDFs of free nucleons and of SRC pairs, with the fraction fully encapsulated in the ratio of neutrons (protons) in SRC pairs $C_{n}^{\rm A}$ ($C_{p}^{\rm A}$). With $R_{\text{HMT}} = 20\%$ in  $^{124}$Sn ($A=124$, $\delta=0.194$), one writes  $C_{n}^{124}=16\%$ and $C_{p}^{124}=24\%$, respectively, in accordance with the results obtained from high energy data \cite{nCTEQ:2023cpo}. The result also underscores the need for further systematic experimental studies of higher-energy  $\gamma$-ray production in HICs.  Interestingly, based on very similar mechanism of the presence of HMT, sub-threshold particle production in proton-nucleus collisions is proposed recently as a probe of SRC in nuclei \cite{Reichert:2025egt}.

{\it Summary $-$} This work presents the first direct and quantitative evidence of short-range correlations (SRCs) in low-energy heavy-ion collisions, based on the highest-quality measurement to date of neutron-proton bremsstrahlung $\gamma$-rays up to 100 MeV in symmetric \snsn~ reactions at 25 MeV/u. This resolves a longstanding puzzle from the 1980s regarding the origin of high-energy $\gamma$-rays in such collisions \cite{NJOCK1986125}, and establishes bremsstrahlung emission as a novel and sensitive probe of SRCs in nuclei. By analyzing the spectral hardening caused by high-momentum nucleons, we determine the SRC fraction in $\rm^{124}Sn$ to be $R_{\rm HMT} = (20 \pm 3)\%$, through comparison with transport model predictions. This result surpasses the previous experimental attempts in both precision and clarity, made possible by improved statistics, extended spectral coverage, and a symmetric reaction system. Our findings open a new  avenue for studying nucleon SRCs and their possible connection to partonic dynamics in nuclei through low-energy heavy-ion collisions.

{\it Acknowledgements.}
This work is supported by the Ministry of Science and Technology under Grant No. 2022YFE0103400, by the National Natural Science Foundation of China under Grant Nos. 12335008, 12205160, 12275129,  by the Center for High Performance Computing and Initiative Scientific Research Program in Tsinghua University and by HIRFL. The authors acknowledge the HIRFL machine team for providing the excellent beam condition.

\bibliography{reference}

\begin{thebibliography}{56}%
\makeatletter
\providecommand \@ifxundefined [1]{%
 \@ifx{#1\undefined}
}%
\providecommand \@ifnum [1]{%
 \ifnum #1\expandafter \@firstoftwo
 \else \expandafter \@secondoftwo
 \fi
}%
\providecommand \@ifx [1]{%
 \ifx #1\expandafter \@firstoftwo
 \else \expandafter \@secondoftwo
 \fi
}%
\providecommand \natexlab [1]{#1}%
\providecommand \enquote  [1]{``#1''}%
\providecommand \bibnamefont  [1]{#1}%
\providecommand \bibfnamefont [1]{#1}%
\providecommand \citenamefont [1]{#1}%
\providecommand \href@noop [0]{\@secondoftwo}%
\providecommand \href [0]{\begingroup \@sanitize@url \@href}%
\providecommand \@href[1]{\@@startlink{#1}\@@href}%
\providecommand \@@href[1]{\endgroup#1\@@endlink}%
\providecommand \@sanitize@url [0]{\catcode `\\12\catcode `\$12\catcode `\&12\catcode `\#12\catcode `\^12\catcode `\_12\catcode `\%12\relax}%
\providecommand \@@startlink[1]{}%
\providecommand \@@endlink[0]{}%
\providecommand \url  [0]{\begingroup\@sanitize@url \@url }%
\providecommand \@url [1]{\endgroup\@href {#1}{\urlprefix }}%
\providecommand \urlprefix  [0]{URL }%
\providecommand \Eprint [0]{\href }%
\providecommand \doibase [0]{https://doi.org/}%
\providecommand \selectlanguage [0]{\@gobble}%
\providecommand \bibinfo  [0]{\@secondoftwo}%
\providecommand \bibfield  [0]{\@secondoftwo}%
\providecommand \translation [1]{[#1]}%
\providecommand \BibitemOpen [0]{}%
\providecommand \bibitemStop [0]{}%
\providecommand \bibitemNoStop [0]{.\EOS\space}%
\providecommand \EOS [0]{\spacefactor3000\relax}%
\providecommand \BibitemShut  [1]{\csname bibitem#1\endcsname}%
\let\auto@bib@innerbib\@empty
%</preamble>
\bibitem [{\citenamefont {Hen}\ \emph {et~al.}(2014)\citenamefont {Hen} \emph {et~al.}}]{Hen:2014nza}%
  \BibitemOpen
  \bibfield  {author} {\bibinfo {author} {\bibfnamefont {O.}~\bibnamefont {Hen}} \emph {et~al.},\ }\bibfield  {title} {\bibinfo {title} {{Momentum sharing in imbalanced Fermi systems}},\ }\href {https://doi.org/10.1126/science.1256785} {\bibfield  {journal} {\bibinfo  {journal} {Science}\ }\textbf {\bibinfo {volume} {346}},\ \bibinfo {pages} {614} (\bibinfo {year} {2014})},\ \Eprint {https://arxiv.org/abs/1412.0138} {arXiv:1412.0138 [nucl-ex]} \BibitemShut {NoStop}%
\bibitem [{\citenamefont {Hen}\ \emph {et~al.}(2017)\citenamefont {Hen}, \citenamefont {Miller}, \citenamefont {Piasetzky},\ and\ \citenamefont {Weinstein}}]{Hen:2016kwk}%
  \BibitemOpen
  \bibfield  {author} {\bibinfo {author} {\bibfnamefont {O.}~\bibnamefont {Hen}}, \bibinfo {author} {\bibfnamefont {G.~A.}\ \bibnamefont {Miller}}, \bibinfo {author} {\bibfnamefont {E.}~\bibnamefont {Piasetzky}},\ and\ \bibinfo {author} {\bibfnamefont {L.~B.}\ \bibnamefont {Weinstein}},\ }\bibfield  {title} {\bibinfo {title} {{Nucleon-Nucleon Correlations, Short-lived Excitations, and the Quarks Within}},\ }\href {https://doi.org/10.1103/RevModPhys.89.045002} {\bibfield  {journal} {\bibinfo  {journal} {Rev. Mod. Phys.}\ }\textbf {\bibinfo {volume} {89}},\ \bibinfo {pages} {045002} (\bibinfo {year} {2017})},\ \Eprint {https://arxiv.org/abs/1611.09748} {arXiv:1611.09748 [nucl-ex]} \BibitemShut {NoStop}%
\bibitem [{\citenamefont {Duer}\ \emph {et~al.}(2018)\citenamefont {Duer} \emph {et~al.}}]{CLAS:2018yvt}%
  \BibitemOpen
  \bibfield  {author} {\bibinfo {author} {\bibfnamefont {M.}~\bibnamefont {Duer}} \emph {et~al.} (\bibinfo {collaboration} {CLAS}),\ }\bibfield  {title} {\bibinfo {title} {{Probing high-momentum protons and neutrons in neutron-rich nuclei}},\ }\href {https://doi.org/10.1038/s41586-018-0400-z} {\bibfield  {journal} {\bibinfo  {journal} {Nature}\ }\textbf {\bibinfo {volume} {560}},\ \bibinfo {pages} {617} (\bibinfo {year} {2018})}\BibitemShut {NoStop}%
\bibitem [{\citenamefont {Xu}\ and\ \citenamefont {Li}(2010)}]{PhysRevC.81.064612}%
  \BibitemOpen
  \bibfield  {author} {\bibinfo {author} {\bibfnamefont {C.}~\bibnamefont {Xu}}\ and\ \bibinfo {author} {\bibfnamefont {B.-A.}\ \bibnamefont {Li}},\ }\bibfield  {title} {\bibinfo {title} {Understanding the major uncertainties in the nuclear symmetry energy at suprasaturation densities},\ }\href {https://doi.org/10.1103/PhysRevC.81.064612} {\bibfield  {journal} {\bibinfo  {journal} {Phys. Rev. C}\ }\textbf {\bibinfo {volume} {81}},\ \bibinfo {pages} {064612} (\bibinfo {year} {2010})}\BibitemShut {NoStop}%
\bibitem [{\citenamefont {Hen}\ \emph {et~al.}(2015)\citenamefont {Hen}, \citenamefont {Li}, \citenamefont {Guo}, \citenamefont {Weinstein},\ and\ \citenamefont {Piasetzky}}]{PhysRevC.91.025803}%
  \BibitemOpen
  \bibfield  {author} {\bibinfo {author} {\bibfnamefont {O.}~\bibnamefont {Hen}}, \bibinfo {author} {\bibfnamefont {B.-A.}\ \bibnamefont {Li}}, \bibinfo {author} {\bibfnamefont {W.-J.}\ \bibnamefont {Guo}}, \bibinfo {author} {\bibfnamefont {L.~B.}\ \bibnamefont {Weinstein}},\ and\ \bibinfo {author} {\bibfnamefont {E.}~\bibnamefont {Piasetzky}},\ }\bibfield  {title} {\bibinfo {title} {Symmetry energy of nucleonic matter with tensor correlations},\ }\href {https://doi.org/10.1103/PhysRevC.91.025803} {\bibfield  {journal} {\bibinfo  {journal} {Phys. Rev. C}\ }\textbf {\bibinfo {volume} {91}},\ \bibinfo {pages} {025803} (\bibinfo {year} {2015})}\BibitemShut {NoStop}%
\bibitem [{\citenamefont {Cai}\ and\ \citenamefont {Li}(2016)}]{PhysRevC.93.014619}%
  \BibitemOpen
  \bibfield  {author} {\bibinfo {author} {\bibfnamefont {B.-J.}\ \bibnamefont {Cai}}\ and\ \bibinfo {author} {\bibfnamefont {B.-A.}\ \bibnamefont {Li}},\ }\bibfield  {title} {\bibinfo {title} {Symmetry energy of cold nucleonic matter within a relativistic mean field model encapsulating effects of high-momentum nucleons induced by short-range correlations},\ }\href {https://doi.org/10.1103/PhysRevC.93.014619} {\bibfield  {journal} {\bibinfo  {journal} {Phys. Rev. C}\ }\textbf {\bibinfo {volume} {93}},\ \bibinfo {pages} {014619} (\bibinfo {year} {2016})}\BibitemShut {NoStop}%
\bibitem [{\citenamefont {{Rittenhouse West}}(2023)}]{RITTENHOUSEWEST2023122563}%
  \BibitemOpen
  \bibfield  {author} {\bibinfo {author} {\bibfnamefont {J.}~\bibnamefont {{Rittenhouse West}}},\ }\bibfield  {title} {\bibinfo {title} {Diquark induced short-range nucleon-nucleon correlations \& the emc effect},\ }\href {https://doi.org/https://doi.org/10.1016/j.nuclphysa.2022.122563} {\bibfield  {journal} {\bibinfo  {journal} {Nuclear Physics A}\ }\textbf {\bibinfo {volume} {1029}},\ \bibinfo {pages} {122563} (\bibinfo {year} {2023})}\BibitemShut {NoStop}%
\bibitem [{\citenamefont {Aubert}\ \emph {et~al.}(1983)\citenamefont {Aubert} \emph {et~al.}}]{EuropeanMuon:1983wih}%
  \BibitemOpen
  \bibfield  {author} {\bibinfo {author} {\bibfnamefont {J.~J.}\ \bibnamefont {Aubert}} \emph {et~al.} (\bibinfo {collaboration} {European Muon}),\ }\bibfield  {title} {\bibinfo {title} {{The ratio of the nucleon structure functions $F2_n$ for iron and deuterium}},\ }\href {https://doi.org/10.1016/0370-2693(83)90437-9} {\bibfield  {journal} {\bibinfo  {journal} {Phys. Lett. B}\ }\textbf {\bibinfo {volume} {123}},\ \bibinfo {pages} {275} (\bibinfo {year} {1983})}\BibitemShut {NoStop}%
\bibitem [{\citenamefont {Weinstein}\ \emph {et~al.}(2011)\citenamefont {Weinstein}, \citenamefont {Piasetzky}, \citenamefont {Higinbotham}, \citenamefont {Gomez}, \citenamefont {Hen},\ and\ \citenamefont {Shneor}}]{Weinstein:2010rt}%
  \BibitemOpen
  \bibfield  {author} {\bibinfo {author} {\bibfnamefont {L.~B.}\ \bibnamefont {Weinstein}}, \bibinfo {author} {\bibfnamefont {E.}~\bibnamefont {Piasetzky}}, \bibinfo {author} {\bibfnamefont {D.~W.}\ \bibnamefont {Higinbotham}}, \bibinfo {author} {\bibfnamefont {J.}~\bibnamefont {Gomez}}, \bibinfo {author} {\bibfnamefont {O.}~\bibnamefont {Hen}},\ and\ \bibinfo {author} {\bibfnamefont {R.}~\bibnamefont {Shneor}},\ }\bibfield  {title} {\bibinfo {title} {{Short Range Correlations and the EMC Effect}},\ }\href {https://doi.org/10.1103/PhysRevLett.106.052301} {\bibfield  {journal} {\bibinfo  {journal} {Phys. Rev. Lett.}\ }\textbf {\bibinfo {volume} {106}},\ \bibinfo {pages} {052301} (\bibinfo {year} {2011})},\ \Eprint {https://arxiv.org/abs/1009.5666} {arXiv:1009.5666 [hep-ph]} \BibitemShut {NoStop}%
\bibitem [{\citenamefont {Hen}\ \emph {et~al.}(2012)\citenamefont {Hen}, \citenamefont {Piasetzky},\ and\ \citenamefont {Weinstein}}]{Hen:2012fm}%
  \BibitemOpen
  \bibfield  {author} {\bibinfo {author} {\bibfnamefont {O.}~\bibnamefont {Hen}}, \bibinfo {author} {\bibfnamefont {E.}~\bibnamefont {Piasetzky}},\ and\ \bibinfo {author} {\bibfnamefont {L.~B.}\ \bibnamefont {Weinstein}},\ }\bibfield  {title} {\bibinfo {title} {{New data strengthen the connection between Short Range Correlations and the EMC effect}},\ }\href {https://doi.org/10.1103/PhysRevC.85.047301} {\bibfield  {journal} {\bibinfo  {journal} {Phys. Rev. C}\ }\textbf {\bibinfo {volume} {85}},\ \bibinfo {pages} {047301} (\bibinfo {year} {2012})},\ \Eprint {https://arxiv.org/abs/1202.3452} {arXiv:1202.3452 [nucl-ex]} \BibitemShut {NoStop}%
\bibitem [{\citenamefont {Schmookler}\ \emph {et~al.}(2019)\citenamefont {Schmookler} \emph {et~al.}}]{CLAS:2019vsb}%
  \BibitemOpen
  \bibfield  {author} {\bibinfo {author} {\bibfnamefont {B.}~\bibnamefont {Schmookler}} \emph {et~al.} (\bibinfo {collaboration} {CLAS}),\ }\bibfield  {title} {\bibinfo {title} {{Modified structure of protons and neutrons in correlated pairs}},\ }\href {https://doi.org/10.1038/s41586-019-0925-9} {\bibfield  {journal} {\bibinfo  {journal} {Nature}\ }\textbf {\bibinfo {volume} {566}},\ \bibinfo {pages} {354} (\bibinfo {year} {2019})},\ \Eprint {https://arxiv.org/abs/2004.12065} {arXiv:2004.12065 [nucl-ex]} \BibitemShut {NoStop}%
\bibitem [{\citenamefont {Norton}(2003)}]{Norton:2003cb}%
  \BibitemOpen
  \bibfield  {author} {\bibinfo {author} {\bibfnamefont {P.~R.}\ \bibnamefont {Norton}},\ }\bibfield  {title} {\bibinfo {title} {{The EMC effect}},\ }\href {https://doi.org/10.1088/0034-4885/66/8/201} {\bibfield  {journal} {\bibinfo  {journal} {Rept. Prog. Phys.}\ }\textbf {\bibinfo {volume} {66}},\ \bibinfo {pages} {1253} (\bibinfo {year} {2003})}\BibitemShut {NoStop}%
\bibitem [{\citenamefont {Geesaman}\ \emph {et~al.}(1995)\citenamefont {Geesaman}, \citenamefont {Saito},\ and\ \citenamefont {Thomas}}]{Geesaman:1995yd}%
  \BibitemOpen
  \bibfield  {author} {\bibinfo {author} {\bibfnamefont {D.~F.}\ \bibnamefont {Geesaman}}, \bibinfo {author} {\bibfnamefont {K.}~\bibnamefont {Saito}},\ and\ \bibinfo {author} {\bibfnamefont {A.~W.}\ \bibnamefont {Thomas}},\ }\bibfield  {title} {\bibinfo {title} {{The nuclear EMC effect}},\ }\href {https://doi.org/10.1146/annurev.ns.45.120195.002005} {\bibfield  {journal} {\bibinfo  {journal} {Ann. Rev. Nucl. Part. Sci.}\ }\textbf {\bibinfo {volume} {45}},\ \bibinfo {pages} {337} (\bibinfo {year} {1995})}\BibitemShut {NoStop}%
\bibitem [{\citenamefont {Malace}\ \emph {et~al.}(2014)\citenamefont {Malace}, \citenamefont {Gaskell}, \citenamefont {Higinbotham},\ and\ \citenamefont {Cloet}}]{Malace:2014uea}%
  \BibitemOpen
  \bibfield  {author} {\bibinfo {author} {\bibfnamefont {S.}~\bibnamefont {Malace}}, \bibinfo {author} {\bibfnamefont {D.}~\bibnamefont {Gaskell}}, \bibinfo {author} {\bibfnamefont {D.~W.}\ \bibnamefont {Higinbotham}},\ and\ \bibinfo {author} {\bibfnamefont {I.}~\bibnamefont {Cloet}},\ }\bibfield  {title} {\bibinfo {title} {{The Challenge of the EMC Effect: existing data and future directions}},\ }\href {https://doi.org/10.1142/S0218301314300136} {\bibfield  {journal} {\bibinfo  {journal} {Int. J. Mod. Phys. E}\ }\textbf {\bibinfo {volume} {23}},\ \bibinfo {pages} {1430013} (\bibinfo {year} {2014})},\ \Eprint {https://arxiv.org/abs/1405.1270} {arXiv:1405.1270 [nucl-ex]} \BibitemShut {NoStop}%
\bibitem [{\citenamefont {Kelly}(1996)}]{Kelly:1996hd}%
  \BibitemOpen
  \bibfield  {author} {\bibinfo {author} {\bibfnamefont {J.~J.}\ \bibnamefont {Kelly}},\ }\bibfield  {title} {\bibinfo {title} {{Nucleon knockout by intermediate-energy electrons}},\ }\href {https://doi.org/10.1007/0-306-47067-5_2} {\bibfield  {journal} {\bibinfo  {journal} {Adv. Nucl. Phys.}\ }\textbf {\bibinfo {volume} {23}},\ \bibinfo {pages} {75} (\bibinfo {year} {1996})}\BibitemShut {NoStop}%
\bibitem [{\citenamefont {Dickhoff}\ and\ \citenamefont {Barbieri}(2004)}]{Dickhoff:2004xx}%
  \BibitemOpen
  \bibfield  {author} {\bibinfo {author} {\bibfnamefont {W.~H.}\ \bibnamefont {Dickhoff}}\ and\ \bibinfo {author} {\bibfnamefont {C.}~\bibnamefont {Barbieri}},\ }\bibfield  {title} {\bibinfo {title} {{Selfconsistent Green's function method for nuclei and nuclear matter}},\ }\href {https://doi.org/10.1016/j.ppnp.2004.02.038} {\bibfield  {journal} {\bibinfo  {journal} {Prog. Part. Nucl. Phys.}\ }\textbf {\bibinfo {volume} {52}},\ \bibinfo {pages} {377} (\bibinfo {year} {2004})},\ \Eprint {https://arxiv.org/abs/nucl-th/0402034} {arXiv:nucl-th/0402034} \BibitemShut {NoStop}%
\bibitem [{\citenamefont {Denniston}\ \emph {et~al.}(2024{\natexlab{a}})\citenamefont {Denniston}, \citenamefont {Je\ifmmode~\check{z}\else \v{z}\fi{}o}, \citenamefont {Kusina}, \citenamefont {Derakhshanian}, \citenamefont {Duwent\"aster}, \citenamefont {Hen}, \citenamefont {Keppel}, \citenamefont {Klasen}, \citenamefont {Kova\ifmmode~\check{r}\else \v{r}\fi{}\'{\i}k}, \citenamefont {Morf\'{\i}n}, \citenamefont {Muzakka}, \citenamefont {Olness}, \citenamefont {Piasetzky}, \citenamefont {Risse}, \citenamefont {Ruiz}, \citenamefont {Schienbein},\ and\ \citenamefont {Yu.}}]{PhysRevLett.133.152502}%
  \BibitemOpen
  \bibfield  {author} {\bibinfo {author} {\bibfnamefont {A.~W.}\ \bibnamefont {Denniston}}, \bibinfo {author} {\bibfnamefont {T.}~\bibnamefont {Je\ifmmode~\check{z}\else \v{z}\fi{}o}}, \bibinfo {author} {\bibfnamefont {A.}~\bibnamefont {Kusina}}, \bibinfo {author} {\bibfnamefont {N.}~\bibnamefont {Derakhshanian}}, \bibinfo {author} {\bibfnamefont {P.}~\bibnamefont {Duwent\"aster}}, \bibinfo {author} {\bibfnamefont {O.}~\bibnamefont {Hen}}, \bibinfo {author} {\bibfnamefont {C.}~\bibnamefont {Keppel}}, \bibinfo {author} {\bibfnamefont {M.}~\bibnamefont {Klasen}}, \bibinfo {author} {\bibfnamefont {K.}~\bibnamefont {Kova\ifmmode~\check{r}\else \v{r}\fi{}\'{\i}k}}, \bibinfo {author} {\bibfnamefont {J.~G.}\ \bibnamefont {Morf\'{\i}n}}, \bibinfo {author} {\bibfnamefont {K.~F.}\ \bibnamefont {Muzakka}}, \bibinfo {author} {\bibfnamefont {F.~I.}\ \bibnamefont {Olness}}, \bibinfo {author} {\bibfnamefont {E.}~\bibnamefont {Piasetzky}}, \bibinfo {author} {\bibfnamefont {P.}~\bibnamefont {Risse}}, \bibinfo {author}
  {\bibfnamefont {R.}~\bibnamefont {Ruiz}}, \bibinfo {author} {\bibfnamefont {I.}~\bibnamefont {Schienbein}},\ and\ \bibinfo {author} {\bibfnamefont {J.~Y.}\ \bibnamefont {Yu.}},\ }\bibfield  {title} {\bibinfo {title} {Modification of quark-gluon distributions in nuclei by correlated nucleon pairs},\ }\href {https://doi.org/10.1103/PhysRevLett.133.152502} {\bibfield  {journal} {\bibinfo  {journal} {Phys. Rev. Lett.}\ }\textbf {\bibinfo {volume} {133}},\ \bibinfo {pages} {152502} (\bibinfo {year} {2024}{\natexlab{a}})}\BibitemShut {NoStop}%
\bibitem [{\citenamefont {Shneor}\ \emph {et~al.}(2007)\citenamefont {Shneor} \emph {et~al.}}]{JeffersonLabHallA:2007lly}%
  \BibitemOpen
  \bibfield  {author} {\bibinfo {author} {\bibfnamefont {R.}~\bibnamefont {Shneor}} \emph {et~al.} (\bibinfo {collaboration} {Jefferson Lab Hall A}),\ }\bibfield  {title} {\bibinfo {title} {{Investigation of proton-proton short-range correlations via the C-12(e, e-prime pp) reaction}},\ }\href {https://doi.org/10.1103/PhysRevLett.99.072501} {\bibfield  {journal} {\bibinfo  {journal} {Phys. Rev. Lett.}\ }\textbf {\bibinfo {volume} {99}},\ \bibinfo {pages} {072501} (\bibinfo {year} {2007})},\ \Eprint {https://arxiv.org/abs/nucl-ex/0703023} {arXiv:nucl-ex/0703023} \BibitemShut {NoStop}%
\bibitem [{\citenamefont {Subedi}\ \emph {et~al.}(2008)\citenamefont {Subedi} \emph {et~al.}}]{Subedi:2008zz}%
  \BibitemOpen
  \bibfield  {author} {\bibinfo {author} {\bibfnamefont {R.}~\bibnamefont {Subedi}} \emph {et~al.},\ }\bibfield  {title} {\bibinfo {title} {{Probing Cold Dense Nuclear Matter}},\ }\href {https://doi.org/10.1126/science.1156675} {\bibfield  {journal} {\bibinfo  {journal} {Science}\ }\textbf {\bibinfo {volume} {320}},\ \bibinfo {pages} {1476} (\bibinfo {year} {2008})},\ \Eprint {https://arxiv.org/abs/0908.1514} {arXiv:0908.1514 [nucl-ex]} \BibitemShut {NoStop}%
\bibitem [{\citenamefont {Korover}\ \emph {et~al.}(2014)\citenamefont {Korover} \emph {et~al.}}]{LabHallA:2014wqo}%
  \BibitemOpen
  \bibfield  {author} {\bibinfo {author} {\bibfnamefont {I.}~\bibnamefont {Korover}} \emph {et~al.} (\bibinfo {collaboration} {Lab Hall A}),\ }\bibfield  {title} {\bibinfo {title} {{Probing the Repulsive Core of the Nucleon-Nucleon Interaction via the $^4He(e,e'pN)$ Triple-Coincidence Reaction}},\ }\href {https://doi.org/10.1103/PhysRevLett.113.022501} {\bibfield  {journal} {\bibinfo  {journal} {Phys. Rev. Lett.}\ }\textbf {\bibinfo {volume} {113}},\ \bibinfo {pages} {022501} (\bibinfo {year} {2014})},\ \Eprint {https://arxiv.org/abs/1401.6138} {arXiv:1401.6138 [nucl-ex]} \BibitemShut {NoStop}%
\bibitem [{\citenamefont {Tang}\ \emph {et~al.}(2003)\citenamefont {Tang} \emph {et~al.}}]{Tang:2002ww}%
  \BibitemOpen
  \bibfield  {author} {\bibinfo {author} {\bibfnamefont {A.}~\bibnamefont {Tang}} \emph {et~al.},\ }\bibfield  {title} {\bibinfo {title} {{n-p short range correlations from (p,2p + n) measurements}},\ }\href {https://doi.org/10.1103/PhysRevLett.90.042301} {\bibfield  {journal} {\bibinfo  {journal} {Phys. Rev. Lett.}\ }\textbf {\bibinfo {volume} {90}},\ \bibinfo {pages} {042301} (\bibinfo {year} {2003})},\ \Eprint {https://arxiv.org/abs/nucl-ex/0206003} {arXiv:nucl-ex/0206003} \BibitemShut {NoStop}%
\bibitem [{\citenamefont {Piasetzky}\ \emph {et~al.}(2006)\citenamefont {Piasetzky}, \citenamefont {Sargsian}, \citenamefont {Frankfurt}, \citenamefont {Strikman},\ and\ \citenamefont {Watson}}]{Piasetzky:2006ai}%
  \BibitemOpen
  \bibfield  {author} {\bibinfo {author} {\bibfnamefont {E.}~\bibnamefont {Piasetzky}}, \bibinfo {author} {\bibfnamefont {M.}~\bibnamefont {Sargsian}}, \bibinfo {author} {\bibfnamefont {L.}~\bibnamefont {Frankfurt}}, \bibinfo {author} {\bibfnamefont {M.}~\bibnamefont {Strikman}},\ and\ \bibinfo {author} {\bibfnamefont {J.~W.}\ \bibnamefont {Watson}},\ }\bibfield  {title} {\bibinfo {title} {{Evidence for the strong dominance of proton-neutron correlations in nuclei}},\ }\href {https://doi.org/10.1103/PhysRevLett.97.162504} {\bibfield  {journal} {\bibinfo  {journal} {Phys. Rev. Lett.}\ }\textbf {\bibinfo {volume} {97}},\ \bibinfo {pages} {162504} (\bibinfo {year} {2006})},\ \Eprint {https://arxiv.org/abs/nucl-th/0604012} {arXiv:nucl-th/0604012} \BibitemShut {NoStop}%
\bibitem [{\citenamefont {Patsyuk}\ \emph {et~al.}(2021)\citenamefont {Patsyuk} \emph {et~al.}}]{Patsyuk2021}%
  \BibitemOpen
  \bibfield  {author} {\bibinfo {author} {\bibfnamefont {M.}~\bibnamefont {Patsyuk}} \emph {et~al.},\ }\bibfield  {title} {\bibinfo {title} {{Unperturbed inverse kinematics nucleon knockout measurements with a 48 GeV/c carbon beam}},\ }\href {https://doi.org/10.1038/s41567-021-01193-4} {\bibfield  {journal} {\bibinfo  {journal} {Nature Phys.}\ }\textbf {\bibinfo {volume} {17}},\ \bibinfo {pages} {693} (\bibinfo {year} {2021})},\ \Eprint {https://arxiv.org/abs/2102.02626} {arXiv:2102.02626 [nucl-ex]} \BibitemShut {NoStop}%
\bibitem [{\citenamefont {Bauer}\ \emph {et~al.}(1986)\citenamefont {Bauer}, \citenamefont {Bertsch}, \citenamefont {Cassing},\ and\ \citenamefont {Mosel}}]{PhysRevC.34.2127}%
  \BibitemOpen
  \bibfield  {author} {\bibinfo {author} {\bibfnamefont {W.}~\bibnamefont {Bauer}}, \bibinfo {author} {\bibfnamefont {G.~F.}\ \bibnamefont {Bertsch}}, \bibinfo {author} {\bibfnamefont {W.}~\bibnamefont {Cassing}},\ and\ \bibinfo {author} {\bibfnamefont {U.}~\bibnamefont {Mosel}},\ }\bibfield  {title} {\bibinfo {title} {Energetic photons from intermediate energy proton- and heavy-ion-induced reactions},\ }\href {https://doi.org/10.1103/PhysRevC.34.2127} {\bibfield  {journal} {\bibinfo  {journal} {Phys. Rev. C}\ }\textbf {\bibinfo {volume} {34}},\ \bibinfo {pages} {2127} (\bibinfo {year} {1986})}\BibitemShut {NoStop}%
\bibitem [{\citenamefont {Martinez}\ \emph {et~al.}(1999)\citenamefont {Martinez} \emph {et~al.}}]{MARTINEZ199928}%
  \BibitemOpen
  \bibfield  {author} {\bibinfo {author} {\bibfnamefont {G.}~\bibnamefont {Martinez}} \emph {et~al.},\ }\bibfield  {title} {\bibinfo {title} {{Photon production in heavy ion collisions close to the pion threshold}},\ }\href {https://doi.org/10.1016/S0370-2693(99)00851-5} {\bibfield  {journal} {\bibinfo  {journal} {Phys. Lett. B}\ }\textbf {\bibinfo {volume} {461}},\ \bibinfo {pages} {28} (\bibinfo {year} {1999})},\ \Eprint {https://arxiv.org/abs/nucl-ex/9910006} {arXiv:nucl-ex/9910006} \BibitemShut {NoStop}%
\bibitem [{\citenamefont {Gan}\ \emph {et~al.}(1994)\citenamefont {Gan}, \citenamefont {Brinkmann}, \citenamefont {Caraley}, \citenamefont {Fineman}, \citenamefont {Kernan}, \citenamefont {McGrath},\ and\ \citenamefont {Danielewicz}}]{Gan1994298}%
  \BibitemOpen
  \bibfield  {author} {\bibinfo {author} {\bibfnamefont {N.}~\bibnamefont {Gan}}, \bibinfo {author} {\bibfnamefont {K.-T.}\ \bibnamefont {Brinkmann}}, \bibinfo {author} {\bibfnamefont {A.}~\bibnamefont {Caraley}}, \bibinfo {author} {\bibfnamefont {B.}~\bibnamefont {Fineman}}, \bibinfo {author} {\bibfnamefont {W.}~\bibnamefont {Kernan}}, \bibinfo {author} {\bibfnamefont {R.}~\bibnamefont {McGrath}},\ and\ \bibinfo {author} {\bibfnamefont {P.}~\bibnamefont {Danielewicz}},\ }\bibfield  {title} {\bibinfo {title} {Neutron-proton bremsstrahlung from low-energy heavy-ion reactions},\ }\href {https://doi.org/10.1103/PhysRevC.49.298} {\bibfield  {journal} {\bibinfo  {journal} {Physical Review C}\ }\textbf {\bibinfo {volume} {49}},\ \bibinfo {pages} {298 } (\bibinfo {year} {1994})}\BibitemShut {NoStop}%
\bibitem [{\citenamefont {Xue}\ \emph {et~al.}(2016)\citenamefont {Xue}, \citenamefont {Xu}, \citenamefont {Yong},\ and\ \citenamefont {Ren}}]{Xue:2016udl}%
  \BibitemOpen
  \bibfield  {author} {\bibinfo {author} {\bibfnamefont {H.}~\bibnamefont {Xue}}, \bibinfo {author} {\bibfnamefont {C.}~\bibnamefont {Xu}}, \bibinfo {author} {\bibfnamefont {G.-C.}\ \bibnamefont {Yong}},\ and\ \bibinfo {author} {\bibfnamefont {Z.}~\bibnamefont {Ren}},\ }\bibfield  {title} {\bibinfo {title} {{Neutron\textendash{}proton bremsstrahlung as a possible probe of high-momentum component in nucleon momentum distribution}},\ }\href {https://doi.org/10.1016/j.physletb.2016.02.044} {\bibfield  {journal} {\bibinfo  {journal} {Phys. Lett. B}\ }\textbf {\bibinfo {volume} {755}},\ \bibinfo {pages} {486} (\bibinfo {year} {2016})}\BibitemShut {NoStop}%
\bibitem [{\citenamefont {Wang}\ \emph {et~al.}(2017)\citenamefont {Wang}, \citenamefont {Xu}, \citenamefont {Ren},\ and\ \citenamefont {Gao}}]{PhysRevC.96.054603}%
  \BibitemOpen
  \bibfield  {author} {\bibinfo {author} {\bibfnamefont {Z.}~\bibnamefont {Wang}}, \bibinfo {author} {\bibfnamefont {C.}~\bibnamefont {Xu}}, \bibinfo {author} {\bibfnamefont {Z.}~\bibnamefont {Ren}},\ and\ \bibinfo {author} {\bibfnamefont {C.}~\bibnamefont {Gao}},\ }\bibfield  {title} {\bibinfo {title} {Probing the high-momentum component in the nucleon momentum distribution by nucleon emission from intermediate-energy nucleus-nucleus collisions},\ }\href {https://doi.org/10.1103/PhysRevC.96.054603} {\bibfield  {journal} {\bibinfo  {journal} {Phys. Rev. C}\ }\textbf {\bibinfo {volume} {96}},\ \bibinfo {pages} {054603} (\bibinfo {year} {2017})}\BibitemShut {NoStop}%
\bibitem [{\citenamefont {{K. Hagel}}(2025)}]{Hagel.WPCF2023}%
  \BibitemOpen
  \bibfield  {author} {\bibinfo {author} {\bibnamefont {{K. Hagel}}},\ }\bibfield  {title} {\bibinfo {title} {Possible signatures of short range correlations in intermediate energy heavy ion collisions},\ }\href {https://doi.org/10.1393/ncc/i2025-25030-5} {\bibfield  {journal} {\bibinfo  {journal} {IL Nuovo Cimento}\ }\textbf {\bibinfo {volume} {48C}},\ \bibinfo {pages} {30} (\bibinfo {year} {2025})}\BibitemShut {NoStop}%
\bibitem [{\citenamefont {Huang}\ \emph {et~al.}(2025)\citenamefont {Huang}, \citenamefont {Meng}, \citenamefont {Pang},\ and\ \citenamefont {Wang}}]{Huang:2025uvc}%
  \BibitemOpen
  \bibfield  {author} {\bibinfo {author} {\bibfnamefont {Y.-J.}\ \bibnamefont {Huang}}, \bibinfo {author} {\bibfnamefont {Z.}~\bibnamefont {Meng}}, \bibinfo {author} {\bibfnamefont {L.-G.}\ \bibnamefont {Pang}},\ and\ \bibinfo {author} {\bibfnamefont {X.-N.}\ \bibnamefont {Wang}},\ }\bibfield  {title} {\bibinfo {title} {{A Novel Deep Learning Method for Detecting Nucleon-Nucleon Correlations}},\ }\href@noop {} {\bibfield  {journal} {\bibinfo  {journal} {arXiv preprint arXiv:2504.00790}\ } (\bibinfo {year} {2025})},\ \bibinfo {note} {arXiv:2504.00790 [nucl-th]}\BibitemShut {NoStop}%
\bibitem [{\citenamefont {Guo}\ \emph {et~al.}(2021)\citenamefont {Guo}, \citenamefont {Li},\ and\ \citenamefont {Yong}}]{Guo:2021zcs}%
  \BibitemOpen
  \bibfield  {author} {\bibinfo {author} {\bibfnamefont {W.-M.}\ \bibnamefont {Guo}}, \bibinfo {author} {\bibfnamefont {B.-A.}\ \bibnamefont {Li}},\ and\ \bibinfo {author} {\bibfnamefont {G.-C.}\ \bibnamefont {Yong}},\ }\bibfield  {title} {\bibinfo {title} {{Imprints of high-momentum nucleons in nuclei on hard photons from heavy-ion collisions near the Fermi energy}},\ }\href {https://doi.org/10.1103/PhysRevC.104.034603} {\bibfield  {journal} {\bibinfo  {journal} {Phys. Rev. C}\ }\textbf {\bibinfo {volume} {104}},\ \bibinfo {pages} {034603} (\bibinfo {year} {2021})},\ \Eprint {https://arxiv.org/abs/2106.08242} {arXiv:2106.08242 [nucl-th]} \BibitemShut {NoStop}%
\bibitem [{\citenamefont {Xu}\ \emph {et~al.}(2013)\citenamefont {Xu}, \citenamefont {Li},\ and\ \citenamefont {Li}}]{Xu:2012hf}%
  \BibitemOpen
  \bibfield  {author} {\bibinfo {author} {\bibfnamefont {C.}~\bibnamefont {Xu}}, \bibinfo {author} {\bibfnamefont {A.}~\bibnamefont {Li}},\ and\ \bibinfo {author} {\bibfnamefont {B.-A.}\ \bibnamefont {Li}},\ }\bibfield  {title} {\bibinfo {title} {{Delineating effects of tensor force on the density dependence of nuclear symmetry energy}},\ }\href {https://doi.org/10.1088/1742-6596/420/1/012090} {\bibfield  {journal} {\bibinfo  {journal} {J. Phys. Conf. Ser.}\ }\textbf {\bibinfo {volume} {420}},\ \bibinfo {pages} {012090} (\bibinfo {year} {2013})},\ \Eprint {https://arxiv.org/abs/1207.1639} {arXiv:1207.1639 [nucl-th]} \BibitemShut {NoStop}%
\bibitem [{\citenamefont {Bertsch}\ and\ \citenamefont {{Das Gupta}}(1988)}]{BERTSCH1988189}%
  \BibitemOpen
  \bibfield  {author} {\bibinfo {author} {\bibfnamefont {G.}~\bibnamefont {Bertsch}}\ and\ \bibinfo {author} {\bibfnamefont {S.}~\bibnamefont {{Das Gupta}}},\ }\bibfield  {title} {\bibinfo {title} {A guide to microscopic models for intermediate energy heavy ion collisions},\ }\href@noop {} {\bibfield  {journal} {\bibinfo  {journal} {Physics Reports}\ }\textbf {\bibinfo {volume} {160}},\ \bibinfo {pages} {189} (\bibinfo {year} {1988})}\BibitemShut {NoStop}%
\bibitem [{\citenamefont {Das}\ \emph {et~al.}(2003{\natexlab{a}})\citenamefont {Das}, \citenamefont {Das~Gupta}, \citenamefont {Gale},\ and\ \citenamefont {Li}}]{PhysRevC.67.034611}%
  \BibitemOpen
  \bibfield  {author} {\bibinfo {author} {\bibfnamefont {C.~B.}\ \bibnamefont {Das}}, \bibinfo {author} {\bibfnamefont {S.}~\bibnamefont {Das~Gupta}}, \bibinfo {author} {\bibfnamefont {C.}~\bibnamefont {Gale}},\ and\ \bibinfo {author} {\bibfnamefont {B.-A.}\ \bibnamefont {Li}},\ }\bibfield  {title} {\bibinfo {title} {Momentum dependence of symmetry potential in asymmetric nuclear matter for transport model calculations},\ }\href {https://doi.org/10.1103/PhysRevC.67.034611} {\bibfield  {journal} {\bibinfo  {journal} {Phys. Rev. C}\ }\textbf {\bibinfo {volume} {67}},\ \bibinfo {pages} {034611} (\bibinfo {year} {2003}{\natexlab{a}})}\BibitemShut {NoStop}%
\bibitem [{\citenamefont {Qin}\ \emph {et~al.}(2024)\citenamefont {Qin} \emph {et~al.}}]{Qin:2023qcn}%
  \BibitemOpen
  \bibfield  {author} {\bibinfo {author} {\bibfnamefont {Y.}~\bibnamefont {Qin}} \emph {et~al.},\ }\bibfield  {title} {\bibinfo {title} {{Probing high-momentum component in nucleon momentum distribution by neutron-proton bremsstrahlung \ensuremath{\gamma}-rays in heavy ion reactions}},\ }\href {https://doi.org/10.1016/j.physletb.2024.138514} {\bibfield  {journal} {\bibinfo  {journal} {Phys. Lett. B}\ }\textbf {\bibinfo {volume} {850}},\ \bibinfo {pages} {138514} (\bibinfo {year} {2024})},\ \Eprint {https://arxiv.org/abs/2307.10717} {arXiv:2307.10717 [nucl-ex]} \BibitemShut {NoStop}%
\bibitem [{\citenamefont {Xu}\ \emph {et~al.}(2024)\citenamefont {Xu}, \citenamefont {Qin}, \citenamefont {Qin}, \citenamefont {Si}, \citenamefont {Zhang}, \citenamefont {Wang}, \citenamefont {Niu}, \citenamefont {Xu},\ and\ \citenamefont {Xiao}}]{Xu:2024oct}%
  \BibitemOpen
  \bibfield  {author} {\bibinfo {author} {\bibfnamefont {J.}~\bibnamefont {Xu}}, \bibinfo {author} {\bibfnamefont {Y.}~\bibnamefont {Qin}}, \bibinfo {author} {\bibfnamefont {Z.}~\bibnamefont {Qin}}, \bibinfo {author} {\bibfnamefont {D.}~\bibnamefont {Si}}, \bibinfo {author} {\bibfnamefont {B.}~\bibnamefont {Zhang}}, \bibinfo {author} {\bibfnamefont {Y.}~\bibnamefont {Wang}}, \bibinfo {author} {\bibfnamefont {Q.}~\bibnamefont {Niu}}, \bibinfo {author} {\bibfnamefont {C.}~\bibnamefont {Xu}},\ and\ \bibinfo {author} {\bibfnamefont {Z.}~\bibnamefont {Xiao}},\ }\bibfield  {title} {\bibinfo {title} {{Reconstruction of Bremsstrahlung \ensuremath{\gamma}-rays spectrum in heavy ion reactions with Richardson-Lucy algorithm}},\ }\href {https://doi.org/10.1016/j.physletb.2024.139009} {\bibfield  {journal} {\bibinfo  {journal} {Phys. Lett. B}\ }\textbf {\bibinfo {volume} {857}},\ \bibinfo {pages} {139009} (\bibinfo {year} {2024})},\ \Eprint {https://arxiv.org/abs/2405.06711} {arXiv:2405.06711 [nucl-th]} \BibitemShut
  {NoStop}%
\bibitem [{\citenamefont {Si}\ \emph {et~al.}(2025{\natexlab{a}})\citenamefont {Si} \emph {et~al.}}]{Si:2024ujh}%
  \BibitemOpen
  \bibfield  {author} {\bibinfo {author} {\bibfnamefont {D.}~\bibnamefont {Si}} \emph {et~al.},\ }\bibfield  {title} {\bibinfo {title} {{The neutron array of the compact spectrometer for heavy ion experiments in Fermi energy region}},\ }\href {https://doi.org/10.1016/j.nima.2024.170055} {\bibfield  {journal} {\bibinfo  {journal} {Nucl. Instrum. Meth. A}\ }\textbf {\bibinfo {volume} {1070}},\ \bibinfo {pages} {170055} (\bibinfo {year} {2025}{\natexlab{a}})},\ \Eprint {https://arxiv.org/abs/2406.18605} {arXiv:2406.18605 [physics.ins-det]} \BibitemShut {NoStop}%
\bibitem [{\citenamefont {Guan}\ \emph {et~al.}(2021)\citenamefont {Guan} \emph {et~al.}}]{Guan:2021tbi}%
  \BibitemOpen
  \bibfield  {author} {\bibinfo {author} {\bibfnamefont {F.}~\bibnamefont {Guan}} \emph {et~al.},\ }\bibfield  {title} {\bibinfo {title} {{A Compact Spectrometer for Heavy Ion Experiments in the Fermi energy regime}},\ }\href {https://doi.org/10.1016/j.nima.2021.165592} {\bibfield  {journal} {\bibinfo  {journal} {Nucl. Instrum. Meth. A}\ }\textbf {\bibinfo {volume} {1011}},\ \bibinfo {pages} {165592} (\bibinfo {year} {2021})}\BibitemShut {NoStop}%
\bibitem [{\citenamefont {Wang}\ \emph {et~al.}(2021)\citenamefont {Wang} \emph {et~al.}}]{Wang:2021jgu}%
  \BibitemOpen
  \bibfield  {author} {\bibinfo {author} {\bibfnamefont {Y.-J.}\ \bibnamefont {Wang}} \emph {et~al.},\ }\bibfield  {title} {\bibinfo {title} {{CSHINE for studies of HBT correlation in Heavy Ion Reactions}},\ }\href {https://doi.org/10.1007/s41365-020-00842-2} {\bibfield  {journal} {\bibinfo  {journal} {Nucl. Sci. Tech.}\ }\textbf {\bibinfo {volume} {32}},\ \bibinfo {pages} {4} (\bibinfo {year} {2021})},\ \Eprint {https://arxiv.org/abs/2101.07352} {arXiv:2101.07352 [physics.ins-det]} \BibitemShut {NoStop}%
\bibitem [{\citenamefont {Guan}\ \emph {et~al.}(2022)\citenamefont {Guan} \emph {et~al.}}]{Guan:2021nfk}%
  \BibitemOpen
  \bibfield  {author} {\bibinfo {author} {\bibfnamefont {F.}~\bibnamefont {Guan}} \emph {et~al.},\ }\bibfield  {title} {\bibinfo {title} {{Track recognition for the \ensuremath{\Delta}E\ensuremath{-}E telescopes with silicon strip detectors}},\ }\href {https://doi.org/10.1016/j.nima.2022.166461} {\bibfield  {journal} {\bibinfo  {journal} {Nucl. Instrum. Meth. A}\ }\textbf {\bibinfo {volume} {1029}},\ \bibinfo {pages} {166461} (\bibinfo {year} {2022})},\ \Eprint {https://arxiv.org/abs/2110.08261} {arXiv:2110.08261 [physics.ins-det]} \BibitemShut {NoStop}%
\bibitem [{\citenamefont {Si}\ \emph {et~al.}(2025{\natexlab{b}})\citenamefont {Si} \emph {et~al.}}]{Si:2025eou}%
  \BibitemOpen
  \bibfield  {author} {\bibinfo {author} {\bibfnamefont {D.}~\bibnamefont {Si}} \emph {et~al.},\ }\bibfield  {title} {\bibinfo {title} {{Extracting Neutron-Neutron Interaction Strength and Spatiotemporal Dynamics of Neutron Emission from the Two-Particle Correlation Function}},\ }\href {https://doi.org/10.1103/PhysRevLett.134.222301} {\bibfield  {journal} {\bibinfo  {journal} {Phys. Rev. Lett.}\ }\textbf {\bibinfo {volume} {134}},\ \bibinfo {pages} {222301} (\bibinfo {year} {2025}{\natexlab{b}})},\ \Eprint {https://arxiv.org/abs/2501.09576} {arXiv:2501.09576 [nucl-ex]} \BibitemShut {NoStop}%
\bibitem [{\citenamefont {Qin}\ \emph {et~al.}(2023)\citenamefont {Qin} \emph {et~al.}}]{Qin:2022mzp}%
  \BibitemOpen
  \bibfield  {author} {\bibinfo {author} {\bibfnamefont {Y.}~\bibnamefont {Qin}} \emph {et~al.},\ }\bibfield  {title} {\bibinfo {title} {{A CsI(Tl) hodoscope on CSHINE for Bremsstrahlung \ensuremath{\gamma}-rays in heavy ion reactions}},\ }\href {https://doi.org/10.1016/j.nima.2023.168330} {\bibfield  {journal} {\bibinfo  {journal} {Nucl. Instrum. Meth. A}\ }\textbf {\bibinfo {volume} {1053}},\ \bibinfo {pages} {168330} (\bibinfo {year} {2023})},\ \Eprint {https://arxiv.org/abs/2212.13498} {arXiv:2212.13498 [physics.ins-det]} \BibitemShut {NoStop}%
\bibitem [{\citenamefont {Xu}\ \emph {et~al.}(2025)\citenamefont {Xu}, \citenamefont {Si}, \citenamefont {Qin}, \citenamefont {Xu}, \citenamefont {Chen}, \citenamefont {Hao}, \citenamefont {Fan}, \citenamefont {Wang},\ and\ \citenamefont {Xiao}}]{Xu:2025erv}%
  \BibitemOpen
  \bibfield  {author} {\bibinfo {author} {\bibfnamefont {J.}~\bibnamefont {Xu}}, \bibinfo {author} {\bibfnamefont {D.}~\bibnamefont {Si}}, \bibinfo {author} {\bibfnamefont {Y.}~\bibnamefont {Qin}}, \bibinfo {author} {\bibfnamefont {M.}~\bibnamefont {Xu}}, \bibinfo {author} {\bibfnamefont {K.}~\bibnamefont {Chen}}, \bibinfo {author} {\bibfnamefont {Z.}~\bibnamefont {Hao}}, \bibinfo {author} {\bibfnamefont {G.}~\bibnamefont {Fan}}, \bibinfo {author} {\bibfnamefont {H.}~\bibnamefont {Wang}},\ and\ \bibinfo {author} {\bibfnamefont {Z.}~\bibnamefont {Xiao}},\ }\bibfield  {title} {\bibinfo {title} {{Linear Response of CsI(Tl) Crystal to Energetic Photons below 20 MeV}},\ }\href@noop {} {\bibfield  {journal} {\bibinfo  {journal} {arXiv preprint arXiv:2503.10098}\ } (\bibinfo {year} {2025})},\ \bibinfo {note} {arXiv:2503.10098 [physics.ins-det]}\BibitemShut {NoStop}%
\bibitem [{\citenamefont {Guo}\ \emph {et~al.}(2022)\citenamefont {Guo} \emph {et~al.}}]{Guo:2022kwc}%
  \BibitemOpen
  \bibfield  {author} {\bibinfo {author} {\bibfnamefont {D.}~\bibnamefont {Guo}} \emph {et~al.},\ }\bibfield  {title} {\bibinfo {title} {{An FPGA-based trigger system for CSHINE}},\ }\href {https://doi.org/10.1007/s41365-022-01149-0} {\bibfield  {journal} {\bibinfo  {journal} {Nucl. Sci. Tech.}\ }\textbf {\bibinfo {volume} {33}},\ \bibinfo {pages} {162} (\bibinfo {year} {2022})},\ \Eprint {https://arxiv.org/abs/2206.15382} {arXiv:2206.15382 [physics.ins-det]} \BibitemShut {NoStop}%
\bibitem [{\citenamefont {Li}\ \emph {et~al.}(1996)\citenamefont {Li}, \citenamefont {Ren}, \citenamefont {Ko},\ and\ \citenamefont {Yennello}}]{Li:1996ix}%
  \BibitemOpen
  \bibfield  {author} {\bibinfo {author} {\bibfnamefont {B.-A.}\ \bibnamefont {Li}}, \bibinfo {author} {\bibfnamefont {Z.-Z.}\ \bibnamefont {Ren}}, \bibinfo {author} {\bibfnamefont {C.~M.}\ \bibnamefont {Ko}},\ and\ \bibinfo {author} {\bibfnamefont {S.~J.}\ \bibnamefont {Yennello}},\ }\bibfield  {title} {\bibinfo {title} {{Isospin dependence of collective flow in heavy ion collisions at intermediate-energies}},\ }\href {https://doi.org/10.1103/PhysRevLett.76.4492} {\bibfield  {journal} {\bibinfo  {journal} {Phys. Rev. Lett.}\ }\textbf {\bibinfo {volume} {76}},\ \bibinfo {pages} {4492} (\bibinfo {year} {1996})},\ \Eprint {https://arxiv.org/abs/nucl-th/9605015} {arXiv:nucl-th/9605015} \BibitemShut {NoStop}%
\bibitem [{\citenamefont {Li}\ \emph {et~al.}(1997)\citenamefont {Li}, \citenamefont {Ko},\ and\ \citenamefont {Ren}}]{Li:1997rc}%
  \BibitemOpen
  \bibfield  {author} {\bibinfo {author} {\bibfnamefont {B.-A.}\ \bibnamefont {Li}}, \bibinfo {author} {\bibfnamefont {C.~M.}\ \bibnamefont {Ko}},\ and\ \bibinfo {author} {\bibfnamefont {Z.-z.}\ \bibnamefont {Ren}},\ }\bibfield  {title} {\bibinfo {title} {{Equation of state of asymmetric nuclear matter and collisions of neutron rich nuclei}},\ }\href {https://doi.org/10.1103/PhysRevLett.78.1644} {\bibfield  {journal} {\bibinfo  {journal} {Phys. Rev. Lett.}\ }\textbf {\bibinfo {volume} {78}},\ \bibinfo {pages} {1644} (\bibinfo {year} {1997})},\ \Eprint {https://arxiv.org/abs/nucl-th/9701048} {arXiv:nucl-th/9701048} \BibitemShut {NoStop}%
\bibitem [{\citenamefont {Li}\ \emph {et~al.}(2004)\citenamefont {Li}, \citenamefont {Das}, \citenamefont {Das~Gupta},\ and\ \citenamefont {Gale}}]{Li:2003ts}%
  \BibitemOpen
  \bibfield  {author} {\bibinfo {author} {\bibfnamefont {B.-A.}\ \bibnamefont {Li}}, \bibinfo {author} {\bibfnamefont {C.~B.}\ \bibnamefont {Das}}, \bibinfo {author} {\bibfnamefont {S.}~\bibnamefont {Das~Gupta}},\ and\ \bibinfo {author} {\bibfnamefont {C.}~\bibnamefont {Gale}},\ }\bibfield  {title} {\bibinfo {title} {{Effects of momentum dependent symmetry potential on heavy ion collisions induced by neutron rich nuclei}},\ }\href {https://doi.org/10.1016/j.nuclphysa.2004.02.016} {\bibfield  {journal} {\bibinfo  {journal} {Nucl. Phys. A}\ }\textbf {\bibinfo {volume} {735}},\ \bibinfo {pages} {563} (\bibinfo {year} {2004})},\ \Eprint {https://arxiv.org/abs/nucl-th/0312054} {arXiv:nucl-th/0312054} \BibitemShut {NoStop}%
\bibitem [{\citenamefont {Li}\ \emph {et~al.}(2018{\natexlab{a}})\citenamefont {Li}, \citenamefont {Cai}, \citenamefont {Chen},\ and\ \citenamefont {Xu}}]{Li:2018lpy}%
  \BibitemOpen
  \bibfield  {author} {\bibinfo {author} {\bibfnamefont {B.-A.}\ \bibnamefont {Li}}, \bibinfo {author} {\bibfnamefont {B.-J.}\ \bibnamefont {Cai}}, \bibinfo {author} {\bibfnamefont {L.-W.}\ \bibnamefont {Chen}},\ and\ \bibinfo {author} {\bibfnamefont {J.}~\bibnamefont {Xu}},\ }\bibfield  {title} {\bibinfo {title} {{Nucleon Effective Masses in Neutron-Rich Matter}},\ }\href {https://doi.org/10.1016/j.ppnp.2018.01.001} {\bibfield  {journal} {\bibinfo  {journal} {Prog. Part. Nucl. Phys.}\ }\textbf {\bibinfo {volume} {99}},\ \bibinfo {pages} {29} (\bibinfo {year} {2018}{\natexlab{a}})},\ \Eprint {https://arxiv.org/abs/1801.01213} {arXiv:1801.01213 [nucl-th]} \BibitemShut {NoStop}%
\bibitem [{\citenamefont {Das}\ \emph {et~al.}(2003{\natexlab{b}})\citenamefont {Das}, \citenamefont {Gupta}, \citenamefont {Gale},\ and\ \citenamefont {Li}}]{Das:2002fr}%
  \BibitemOpen
  \bibfield  {author} {\bibinfo {author} {\bibfnamefont {C.~B.}\ \bibnamefont {Das}}, \bibinfo {author} {\bibfnamefont {S.~D.}\ \bibnamefont {Gupta}}, \bibinfo {author} {\bibfnamefont {C.}~\bibnamefont {Gale}},\ and\ \bibinfo {author} {\bibfnamefont {B.-A.}\ \bibnamefont {Li}},\ }\bibfield  {title} {\bibinfo {title} {{Momentum dependence of symmetry potential in asymmetric nuclear matter for transport model calculations}},\ }\href {https://doi.org/10.1103/PhysRevC.67.034611} {\bibfield  {journal} {\bibinfo  {journal} {Phys. Rev. C}\ }\textbf {\bibinfo {volume} {67}},\ \bibinfo {pages} {034611} (\bibinfo {year} {2003}{\natexlab{b}})},\ \Eprint {https://arxiv.org/abs/nucl-th/0212090} {arXiv:nucl-th/0212090} \BibitemShut {NoStop}%
\bibitem [{\citenamefont {Li}\ \emph {et~al.}(2018{\natexlab{b}})\citenamefont {Li}, \citenamefont {Cai}, \citenamefont {Chen},\ and\ \citenamefont {Xu}}]{LI201829}%
  \BibitemOpen
  \bibfield  {author} {\bibinfo {author} {\bibfnamefont {B.-A.}\ \bibnamefont {Li}}, \bibinfo {author} {\bibfnamefont {B.-J.}\ \bibnamefont {Cai}}, \bibinfo {author} {\bibfnamefont {L.-W.}\ \bibnamefont {Chen}},\ and\ \bibinfo {author} {\bibfnamefont {J.}~\bibnamefont {Xu}},\ }\bibfield  {title} {\bibinfo {title} {Nucleon effective masses in neutron-rich matter},\ }\href {https://doi.org/https://doi.org/10.1016/j.ppnp.2018.01.001} {\bibfield  {journal} {\bibinfo  {journal} {Progress in Particle and Nuclear Physics}\ }\textbf {\bibinfo {volume} {99}},\ \bibinfo {pages} {29} (\bibinfo {year} {2018}{\natexlab{b}})}\BibitemShut {NoStop}%
\bibitem [{\citenamefont {Rios}\ \emph {et~al.}(2009)\citenamefont {Rios}, \citenamefont {Polls},\ and\ \citenamefont {Dickhoff}}]{PhysRevC.79.064308}%
  \BibitemOpen
  \bibfield  {author} {\bibinfo {author} {\bibfnamefont {A.}~\bibnamefont {Rios}}, \bibinfo {author} {\bibfnamefont {A.}~\bibnamefont {Polls}},\ and\ \bibinfo {author} {\bibfnamefont {W.~H.}\ \bibnamefont {Dickhoff}},\ }\bibfield  {title} {\bibinfo {title} {Depletion of the nuclear fermi sea},\ }\href {https://doi.org/10.1103/PhysRevC.79.064308} {\bibfield  {journal} {\bibinfo  {journal} {Phys. Rev. C}\ }\textbf {\bibinfo {volume} {79}},\ \bibinfo {pages} {064308} (\bibinfo {year} {2009})}\BibitemShut {NoStop}%
\bibitem [{\citenamefont {Yin}\ \emph {et~al.}(2013)\citenamefont {Yin}, \citenamefont {Li}, \citenamefont {Wang},\ and\ \citenamefont {Zuo}}]{PhysRevC.87.014314}%
  \BibitemOpen
  \bibfield  {author} {\bibinfo {author} {\bibfnamefont {P.}~\bibnamefont {Yin}}, \bibinfo {author} {\bibfnamefont {J.-Y.}\ \bibnamefont {Li}}, \bibinfo {author} {\bibfnamefont {P.}~\bibnamefont {Wang}},\ and\ \bibinfo {author} {\bibfnamefont {W.}~\bibnamefont {Zuo}},\ }\bibfield  {title} {\bibinfo {title} {Three-body force effect on nucleon momentum distributions in asymmetric nuclear matter within the framework of the extended brueckner-hartree-fock approach},\ }\href {https://doi.org/10.1103/PhysRevC.87.014314} {\bibfield  {journal} {\bibinfo  {journal} {Phys. Rev. C}\ }\textbf {\bibinfo {volume} {87}},\ \bibinfo {pages} {014314} (\bibinfo {year} {2013})}\BibitemShut {NoStop}%
\bibitem [{\citenamefont {Wolter}\ \emph {et~al.}(2022)\citenamefont {Wolter} \emph {et~al.}}]{TMEP:2022xjg}%
  \BibitemOpen
  \bibfield  {author} {\bibinfo {author} {\bibfnamefont {H.}~\bibnamefont {Wolter}} \emph {et~al.} (\bibinfo {collaboration} {TMEP}),\ }\bibfield  {title} {\bibinfo {title} {{Transport model comparison studies of intermediate-energy heavy-ion collisions}},\ }\href {https://doi.org/10.1016/j.ppnp.2022.103962} {\bibfield  {journal} {\bibinfo  {journal} {Prog. Part. Nucl. Phys.}\ }\textbf {\bibinfo {volume} {125}},\ \bibinfo {pages} {103962} (\bibinfo {year} {2022})},\ \Eprint {https://arxiv.org/abs/2202.06672} {arXiv:2202.06672 [nucl-th]} \BibitemShut {NoStop}%
\bibitem [{\citenamefont {Denniston}\ \emph {et~al.}(2024{\natexlab{b}})\citenamefont {Denniston} \emph {et~al.}}]{nCTEQ:2023cpo}%
  \BibitemOpen
  \bibfield  {author} {\bibinfo {author} {\bibfnamefont {A.~W.}\ \bibnamefont {Denniston}} \emph {et~al.} (\bibinfo {collaboration} {nCTEQ}),\ }\bibfield  {title} {\bibinfo {title} {{Modification of Quark-Gluon Distributions in Nuclei by Correlated Nucleon Pairs}},\ }\href {https://doi.org/10.1103/PhysRevLett.133.152502} {\bibfield  {journal} {\bibinfo  {journal} {Phys. Rev. Lett.}\ }\textbf {\bibinfo {volume} {133}},\ \bibinfo {pages} {152502} (\bibinfo {year} {2024}{\natexlab{b}})},\ \Eprint {https://arxiv.org/abs/2312.16293} {arXiv:2312.16293 [hep-ph]} \BibitemShut {NoStop}%
\bibitem [{\citenamefont {Reichert}\ and\ \citenamefont {Aichelin}(2025)}]{Reichert:2025egt}%
  \BibitemOpen
  \bibfield  {author} {\bibinfo {author} {\bibfnamefont {T.}~\bibnamefont {Reichert}}\ and\ \bibinfo {author} {\bibfnamefont {J.}~\bibnamefont {Aichelin}},\ }\bibfield  {title} {\bibinfo {title} {{The influence of nuclear short range correlations on sub-threshold particle production in proton-nucleus collisions}},\ }\href@noop {} {\bibfield  {journal} {\bibinfo  {journal} {arXiv preprint arXiv:2506.03962}\ } (\bibinfo {year} {2025})},\ \bibinfo {note} {arXiv:2506.03962 [nucl-th]}\BibitemShut {NoStop}%
\bibitem [{\citenamefont {Njock}\ \emph {et~al.}(1986)\citenamefont {Njock}, \citenamefont {Maurel}, \citenamefont {Monnand}, \citenamefont {Nifenecker}, \citenamefont {Pinston}, \citenamefont {Schussler},\ and\ \citenamefont {Barneoud}}]{NJOCK1986125}%
  \BibitemOpen
  \bibfield  {author} {\bibinfo {author} {\bibfnamefont {M.}~\bibnamefont {Njock}}, \bibinfo {author} {\bibfnamefont {M.}~\bibnamefont {Maurel}}, \bibinfo {author} {\bibfnamefont {E.}~\bibnamefont {Monnand}}, \bibinfo {author} {\bibfnamefont {H.}~\bibnamefont {Nifenecker}}, \bibinfo {author} {\bibfnamefont {J.}~\bibnamefont {Pinston}}, \bibinfo {author} {\bibfnamefont {F.}~\bibnamefont {Schussler}},\ and\ \bibinfo {author} {\bibfnamefont {D.}~\bibnamefont {Barneoud}},\ }\bibfield  {title} {\bibinfo {title} {High energy gamma-ray production in heavy-ion reactions},\ }\href {https://doi.org/https://doi.org/10.1016/0370-2693(86)90700-8} {\bibfield  {journal} {\bibinfo  {journal} {Physics Letters B}\ }\textbf {\bibinfo {volume} {175}},\ \bibinfo {pages} {125} (\bibinfo {year} {1986})}\BibitemShut {NoStop}%
\end{thebibliography}%

\end{document}